\documentclass[final]{elsarticle}
\usepackage{siunitx}
\usepackage{pgfplots}
\usepackage{lineno,hyperref}
\usepackage[]{algorithm2e}
\biboptions{sort&compress}
\usepackage{xcolor}
\definecolor{almond}{rgb}{0.94, 0.87, 0.8}
\definecolor{chamoisee}{rgb}{0.63, 0.47, 0.35}

\modulolinenumbers[5]
\usepackage{fixmath}
\usepackage{mathtools}
\usepackage{amsmath}
\usepackage{tabularx}
\usepackage{tikz}
\usetikzlibrary[patterns]
\usetikzlibrary{arrows,positioning,shapes} 
\usetikzlibrary{intersections, pgfplots.fillbetween}
\usepackage{float}
\usepackage{mwe}
\usepackage{tikzscale}
\usepackage{subcaption} 
\usepackage{gensymb}
\usetikzlibrary{calc} 
\usetikzlibrary{shapes,arrows,positioning,shadows,trees}
\usetikzlibrary{decorations.pathreplacing}
\usetikzlibrary{positioning}
\usetikzlibrary{shapes.misc}
\usetikzlibrary{patterns}
\usetikzlibrary{arrows,positioning,shapes} 
\usepackage[toc,page]{appendix}
\usepackage{amssymb}
\usepackage[export]{adjustbox}
\usepackage{graphicx}
\usepackage{xcolor,colortbl}
\usepackage{tkz-graph}
\usetikzlibrary{external}
\tikzsetexternalprefix{PAPER_ERROR_AND_LOSS/FIGURES_MOSTAFA_PDF/}
\usepackage{listings}

\pgfplotsset{grid style={solid,white!90!black}}

\journal{Journal of Computational Science}

\newcommand{\neww}{\color{black}}

\makeatletter
\def\ps@pprintTitle{%
	\let\@oddhead\@empty
	\let\@evenhead\@empty
	\def\@oddfoot{\reset@font\hfil\thepage\hfil}
	\let\@evenfoot\@oddfoot
}
\makeatother

\usepackage{morefloats}
\RequirePackage{snapshot}

\begin{document}
	
	\begin{frontmatter}
		\title{Error Control and Loss Functions for the Deep Learning Inversion of Borehole Resistivity Measurements}
		
		\author
		{M. Shahriari$^{1,2}$, D. Pardo$^{3,4,5}$, J. A. Rivera$^{3,4}$, C. Torres-Verd\'in$^{6}$, \\  A. Picon$^{7,3}$ J. Del Ser$^{7,3,4}$,  S. Ossand\'on$^{8}$, V. M. Calo$^{9}$\\
			{\footnotesize $^{1}$ Software Competence Center Hagenberg (SCCH), Hagenberg, Austria.\\ 
				$^{2}$ Euskampus Fundazioa, Bilbao, Spain.\\ 
				$^{3}$ University of the Basque Country (UPV/EHU), Leioa, Spain.\\
				$^{4}$ Basque Center for Applied Mathematics, (BCAM), Bilbao, Spain.\\
				$^{5}$ Ikerbasque (Basque Foundation for Sciences), Bilbao, Spain.\\
				$^{6}$ The University of Texas at Austin, USA.\\
				$^{7}$ Tecnalia, Basque Research \& Technology Alliance, 48170 Derio, Spain.\\
				$^{8}$ Pontificia Universidad Cat\'olica de Valpara\'iso, Valpara\'iso, Chile.\\
				$^{9}$ Curtin University, Perth, Australia.}
		}
		
	\begin{abstract}
		Deep learning (DL) is a numerical method that approximates functions. Recently, its use has become attractive for the simulation and inversion of multiple problems in computational mechanics, including the inversion of borehole logging measurements for oil and gas applications. In this context, DL methods exhibit two key attractive features: a) once trained, they enable to solve an inverse problem in a fraction of a second, which is convenient for borehole geosteering operations as well as in other real-time inversion applications. b) DL methods exhibit a superior capability for approximating highly-complex functions across different areas of knowledge. Nevertheless, as it occurs with most numerical methods, DL also relies on expert design decisions that are problem specific to achieve reliable and robust results. Herein, we investigate two key aspects of deep neural networks (DNNs) when applied to the inversion of borehole resistivity measurements: error control and adequate selection of the loss function. As we illustrate via theoretical considerations and extensive numerical experiments, these interrelated aspects are critical to recover accurate inversion results.  

	\end{abstract}
	
		
	\end{frontmatter}
	
	\section{Introduction}
In the last decade, deep learning (DL) algorithms have appealed to the masses due to their high performance in different applications, such as computer vision~\cite{Lu}, speech recognition~\cite{Yu}, and biometrics~\cite{Kumar}, to mention a few. In recent years, there have been significant advances in the field of DL, with the appearance of residual neural networks (RNNs)~\cite{He}, which prevent gradient degeneration during the training stage, and Encoder-Decoder (sequence-to-sequence) deep neural networks (DNNs), which have improved the DL work capability in computer vision applications~\cite{Badrinarayanan}. Due to the high demand for DNNs from industry, dedicated libraries and packages such as Tensorflow~\cite{Lu}, Keras~\cite{chollet2015}, and Pytorch~\cite{paszke2017automatic} have been developed. These libraries facilitate the use of DNNs across different industrial applications~\cite{ZHAO2019213,6639345,LANGKVIST201411,quteprints127354,JAN2019275}. All these advances combined make DNNs one of the most powerful and fast-growing artificial intelligence (AI) tools presently. 

In this work, we focus on the application of DNNs to geosteering operations~\cite{Shahriari_deep_inverse,Pardo,Ijasan}. In this oil \& gas application, a logging-while-drilling (LWD) instrument records electromagnetic measurements, which are inverted in real time to produce a map of the Earth's subsurface. Based on the reconstructed Earth model, the operator adjusts the well-trajectory in real time to further explore exploitation targets, including oil \& gas reservoirs, and to maximize the posterior productivity of the available reserves. Due to the tremendous productivity increase achieved with this technique, nowadays geosteering plays an essential role in the oil \& gas industry~\cite{Desbrandes}. 

The main difficulty one faces when dealing with geosteering problems is the real-time adjustment of the well trajectory. For that, we require a rapid inversion technique. Unfortunately, traditional inversion methods have severe limitations, which force geophysicists to continuously look for new solutions to this problem (see, e.g.,~\cite{Ghasemi,Chemali, Pardo,Dupuis, Zhang,Seifert, Shahriari_deep_inverse, Ijasan}). Gradient-based methods require simulating the forward problem dozens of times for each set of measurements. Moreover, these methods also estimate the derivatives of the measurements with respect to the inversion variables, which is challenging and time consuming~\cite{Tarantola}. To alleviate the high computational costs associated with this inversion method, simplified 1.5-dimensional (1.5D) methods are common (see, e.g.,~\cite{Shahriari, Pardo, Ijasan}). For the inversion of borehole resistivity measurements, an alternative is to apply statistics-based methods~\cite{8746808,Malinverno_2000,doi:10.1190/1.2713043}. The statistical methods perform forward simulations hundreds of times, which also require large computation times~\cite{Vogel}. Both gradient and statistics-based methods \textit{only evaluate} the inverse operator. Thus, the entire inversion process is repeated at each new logging position. 

Below, we employ DNNs to approximate the inverse operator. Although the training stage of a DNN may be time consuming, after the network is properly trained, it can forecast in a fraction of a second~\cite{Shahriari_deep_inverse}. This rapid inversion facilitates geosteering operations. 

DNNs also face important challenges when applied to the inversion of borehole resistivity problems. In particular, to properly train a DNN,  we require a large dataset (also known as \textit{ground truth}) with the solution of the forward problem for different Earth models~\cite{Higham, Shahriari_deep_inverse, Shahriari_deep_forward}. Building a dataset may be time consuming, especially for two and three-dimensional problems. In those cases, we need to solve the forward problem using numerical simulation methods such as the finite element (FEM)~\cite{Shahriari, Aramberri, Bakr} or finite difference (FDM)~\cite{Davydycheva, Davydycheva1}. Moreover, we need to optimally sample the parameter space describing relevant Earth models. Additionally, the training stage can be time consuming. However, this is an \textit{offline} cost. One additional challenge arises due to the nature of inverse problems: they are not well-defined, that is, there may exist multiple outputs for a given input~\cite{Tarantola, Vogel}. As we shall illustrate in this work, when using a DNN equipped with a traditional loss function based on the data misfit, the corresponding DNN approximations may be far away from any of the existing solutions to the inverse operator. This can seriously compromise the reliability of the method and, consequently, the corresponding decision-making during geosteering operations.
	
In this work, we investigate the selection of the loss function to train a DNN when dealing with an inverse problem. We also introduce some error control during training. We focus on the inversion of borehole resistivity measurements. Nonetheless, most of the design decisions of such loss function are applicable to other inverse problems. To explain the main results stemming from this work, we first illustrate them with a simple mathematical example. Then, we apply the resulting DNN approximations to synthetic examples, which help us elucidate their main advantages and limitations. This work does not discuss optimal data sampling techniques nor the decision-making for the optimal selection of DNN architectures \cite{Puzyrev,Moghadas}. Those subjects are possible future work. However, for this article to be self-contained, we briefly describe in the appendix the architecture of the DNN we use.

The remainder of this article is organized as follows. Section~\ref{problem_formulation} states the problem formulation and introduces two examples. In the first one, the exact solution of the inverse problem is the square root of the input data. This example serves to illustrate some of the main features and limitations of DL algorithms. The second example reproduces a realistic inversion scenario of borehole resistivity measurements. We also describe the selected parameterization of the Earth models. Section~\ref{data_space_and_GT} describes the finite-dimensional input and output vector representations of the inverse operator that we approximate via DL. This section also discusses how we generate the ground truth dataset. Section~\ref{data_preprocessing} proposes a preprocessing of the input and output data variables to ensure that contributions to the loss (cost) function corresponding to different measurements and Earth parameters are comparable in magnitude. Section~\ref{norms_and_errors} describes the vector and matrix norms employed in this work, along with the corresponding absolute and relative errors. Section~\ref{loss_function} analyzes various loss functions and illustrates their most prominent advantages and limitations. Section~\ref{implementation} describes the main implementation aspects of our DL inversion algorithm. We present several numerical inversion results of borehole logging measurements in Section~\ref{sec:numerical_results}. In addition to some conclusions and future work we describe in Section~\ref{conclusions}, the manuscript also contains two appendices. Appendix~\ref{app:meas} describes the borehole measurement acquisition system, including the employed logging instruments and recorded measurements. Appendix~\ref{sec:dnn_arquitectures} details the selected DNN architectures.
	\	\section{Problem Formulation} \label{problem_formulation}
	\subsection{Forward problem}
	We fix the measurement acquisition system $\tilde{\mathbf{s}}$. Then, for a well trajectory $\tilde{\mathbf{t}}$, and an Earth model $\tilde{\mathbf{p}}$, the forward problem consists of finding the corresponding borehole resistivity measurements $\tilde{\mathbf{m}}$. We denote by $\tilde{\mathcal{F}}$ the associated forward function. That is:
		\begin{equation}
	\label{forward_function}
	\tilde{\mathcal{F}}(\tilde{\mathbf{t}},\tilde{\mathbf{p}}) = \tilde{\mathbf{m}}, \qquad\hbox{where } \tilde{\mathbf{t}}\in\tilde{\mathbb{T}},\tilde{\mathbf{p}}\in\tilde{\mathbb{P}},\tilde{\mathbf{m}}\in\tilde{\mathbb{M}}.
	\end{equation}
Above, we omit, for convenience, the explicit dependence of the function $\tilde{\mathcal{F}}$ upon the fixed input variable $\tilde{\mathbf{s}}$. $\tilde{\mathbb{P}}$ and $\tilde{\mathbb{M}}$ are normed vector spaces equipped with norms $||\cdot||_{\tilde{\mathbb{P}}}$ and $||\cdot||_{\tilde{\mathbb{M}}}$ , respectively. $\tilde{\mathbb{T}}$ is also a vector space. Function $\tilde{\mathcal{F}}$ consists of a boundary value problem governed by Maxwell's equations (see~\cite{Shahriari} for details).

	\subsection{Inverse problem}\label{InvPro}
	In the inversion of borehole resistivity measurements, the objective is to determine the subsurface properties $\tilde{\mathbf{p}}$ corresponding to a set of measurements $\tilde{\mathbf{m}}$ recorded over a given trajectory $\tilde{{\bf t}}$. Again, the measurement acquisition system $\tilde{\mathbf{s}}$ is fixed. We denote that inverse operator as $\tilde{\mathcal{I}}$ (inverse of $\tilde{\mathcal{F}}$). Mathematically, we have:
	\begin{equation}
		\label{inverse_function}
	\tilde{\mathcal{I}}(\tilde{\mathbf{t}},\tilde{\mathbf{m}})= \tilde{\mathbf{p}}, \qquad\hbox{where } \tilde{\mathbf{t}}\in\tilde{\mathbb{T}},\tilde{\mathbf{m}}\in\tilde{\mathbb{M}},\;\tilde{\mathbf{p}}\in\tilde{\mathbb{P}}.
		\end{equation}
		Again, we omit for convenience the explicit dependence of function $\tilde{\mathcal{I}}$ upon input variable $\tilde{\mathbf{s}}$. The governing physical equation of operator $\tilde{\mathcal{I}}$ is unknown. However, we know that a given input may have multiple associated outputs. Thus, such inverse operator is not well-defined.
		
\begin{figure}[!h]
		\begin{tikzpicture}[scale=1.0]
		\node (layers) at (0,0)[scale=0.75]{
			\input{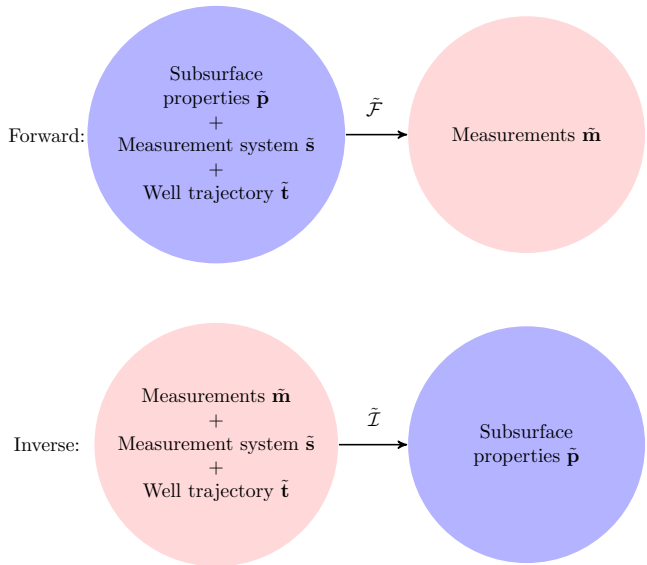} 
		};
		\end{tikzpicture}
		\caption{High-level description of a forward and an inverse problem.}
		\label{problem_types}
	\end{figure}

		Figure~\ref{problem_types} illustrates both forward and inverse problems.
	
	\subsection{Parameterization}\label{Parameterization}
	We select a finite dimensional subspace of $\tilde{\mathbb{T}}$ parameterized with $n_t$ real-valued numbers. The corresponding vector representation of an element from that subspace is ${\bf t} \in \mathbb{R}^{n_t}$. We similarly parameterize a finite dimensional subspace of $\tilde{\mathbb{P}}$ and $\tilde{\mathbb{M}}$ with $n_p$ and $n_m$ real-valued numbers, respectively. The corresponding vector representations of an element from those subspaces are ${\bf p} \in \mathbb{R}^{n_p}$ and ${\bf m} \in \mathbb{R}^{n_m}$, respectively.  
	
	       The span of vector representations ${\bf p}$ and ${\bf m}$ constitute two subspaces of $\mathbb{R}^{n_p}$ and $\mathbb{R}^{n_m}$ with norms $||\cdot||_\mathbb{P}$ and $||\cdot||_\mathbb{M}$, respectively. Ideally, these norms should be inherited from those associated with the original infinite dimensional spaces. However, this is often a challenging task and an open area of research. We directly employ some existing (typically $l_1$ or $l_2$) finite dimensional norms. 
	       
	       The function $\mathcal{F}$ associates a pair ($\mathbf{t}$, $\mathbf{p}$) (vector representations of $(\tilde{\mathbf{t}}$,$\tilde{\mathbf{p}}$)) with $\mathbf{m}$ (vector representation of $\tilde{\mathbf{m}}$) such that $\mathcal{F}(\mathbf{t},\mathbf{p})=\mathbf{m}$. We employ a similar notation for its inverse $\mathcal{I}$ acting on vector representations.
	   
	   To provide context and guidance for future developments, we introduce simple examples that illustrate some of the shortcomings of the standard techniques when applied to these problems, and we explain how we seek to overcome the associated challenges. The first problem seeks to predict the inverse of squaring a number. The second example focuses on geosteering applications.
	       
	\subsection{Example A: Model problem with known analytical solution}\label{Ex:model_problem}
\begin{figure}[!htp]
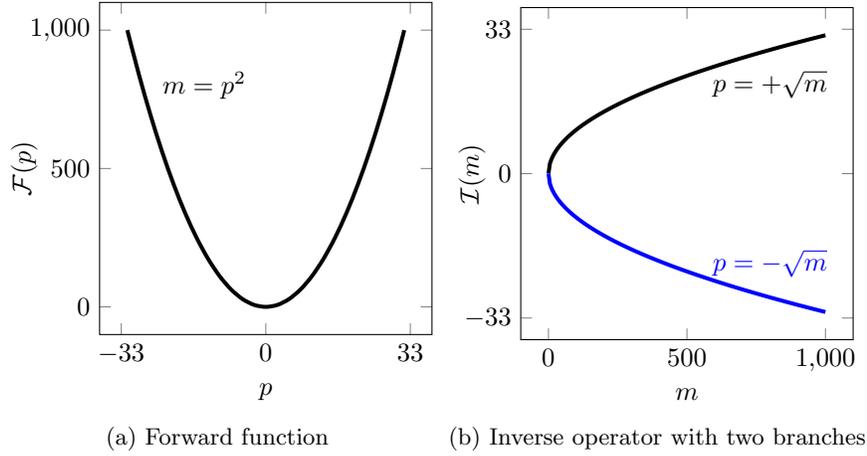

	\subcaptionbox{Forward function}{%
	\input{Jon/Fig_example1.tex}} 
	\subcaptionbox{Inverse operator with two branches}{%
	\input{Jon/Fig_example2.tex}}
	\caption{Model problem with known analytical solution.}
	\label{fig:real_graphics}
\end{figure}

	We select $n_t=0$, $n_p=n_m=1$. The forward function is given by $\mathcal{F}(p)=p^2$, while the inverse has two solutions (branches): $\mathcal{I}(m)=+\sqrt{m}$, and $\mathcal{I}(m)=-\sqrt{m}$, as described in Figure~\ref{fig:real_graphics}.

	This simple example contains a key feature exhibited by most inverse problems: it has multiple solutions. Thus, it illustrates the behaviour of DNNs when considering different loss functions. Results are enlightening and, as we show below, they provide clear guidelines to construct proper loss functions for approximating inverse problems. 
		
	\subsection{Example B: Inversion of borehole resistivity measurements}\label{Ex:borehole}
	In geosteering applications, multiple oil and service companies perform inversion assuming a piecewise 1D layered model of the Earth. In this case, there exist semi-analytic methods that can simulate the forward problem in a fraction of a second. Herein, we use the same approach. Thus, the evaluation of $\mathcal{F}$ is performed with a 1.5D semi-analytic code (see~\cite{Shahriari, Loseth}). As a result, at each logging position, our inversion operator recovers the formation properties of a 1D layered medium~\cite{Pardo, Ijasan}. 
		
\begin{figure}[!h]
		\centering
		\begin{tikzpicture}[scale=1.0]
		\node (layers) at (0,0)[scale=1]{
			\input{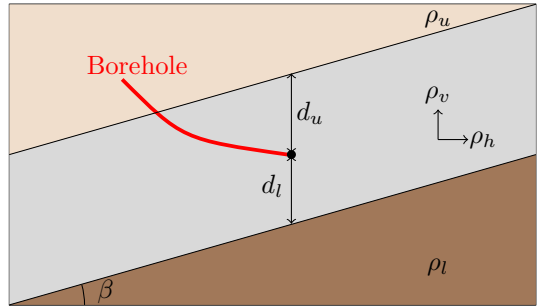} 
		};
		\end{tikzpicture}
		\caption{1D media and a well trajectory. The black circle indicates the last trajectory position.}
		\label{fig_1D}
	\end{figure}

\begin{figure}[!h]
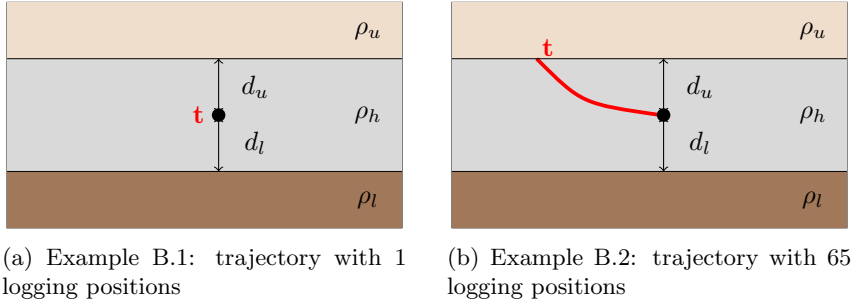

	\centering
	\subcaptionbox{Example B.1: trajectory with 1 logging positions}{%
		\input{Model/1D_1_posittion.tex}}
	\hspace{0.3cm}
	\subcaptionbox{Example B.2: trajectory with 65 logging positions}{%
		\input{Model/1D_65_posittion.tex}}
	\caption{Model problems corresponding to examples B.1 and B.2, respectively.}
	\label{fig_1Dv2}
\end{figure}

	 For our borehole resistivity applications, we assume the Earth model to be a three-layer medium, as Figure~\ref{fig_1D} illustrates. A common practice in the field is to characterize this medium by seven parameters, namely: (a) $\rho_{h}$ and $\rho_{v}$, the horizontal and vertical resistivities of the host layer, respectively; (b) $\rho_{u}$ and $\rho_{l}$, the resistivities of the upper and lower isotropic layers, respectively; (c) $d_{u}$ and $d_{l}$, the distances from the logging position to the upper and lower bed boundaries, respectively; and (d) $\beta$, the dipping angle of the formation. In this work, to simplify the problem, we consider only five of them by restricting the search to isotropic formations ($\rho_{v}=\rho_h$) with zero dip angle ($\beta=0$). Thus, $n_p=5$.
   In this example, we consider two cases (see Figure~\ref{fig_1Dv2}) according with different numbers of logging positions per data sample.
	\subsubsection{Example B.1: one logging position}
For each sample, the input data are the acquired measurements at a single logging position. In this case, $n_m=6$. 
   	\subsubsection{Example B.2: 65 logging positions}
For each sample, we select as input measurements those corresponding to a full logging trajectory formed by 65 logging positions. Thus, for each Earth model $\mathbf{p}$, we parameterize $\mathbf{m}$ with 390 real numbers (i.e., $n_m=390$).

\textbf{Remark:} Appendix~\ref{app:meas} describes the logging instruments, positions, and post-processing system employed to record the measurements.

	\section{Data Space and Ground Truth} \label{data_space_and_GT}

In this work, we employ a deep neural network (DNN) to approximate the discrete inverse operator $\mathcal{I}$. Given a supervised database of $n$-pairs $(\mathbf{m}_i, \mathcal{I}(\mathbf{t}_i,\mathbf{m}_i))$, $i=1,...,n$, the DNN builds an approximation of the unknown function $\mathcal{I}$. This section describes the construction of the supervised database.

We first select the number of samples, $n$, and two subspaces of $\mathbb{R}^{n_p}$ and $\mathbb{R}^{n_t}$, respectively. Then, we select the $n$ samples in those subspaces, namely, $((\mathbf{t}_1,\mathbf{p}_1),...,(\mathbf{t}_n,\mathbf{p}_n))$. To each of these samples, we apply the operator $\mathcal{F}$. That is, we compute $(\mathcal{F}(\mathbf{t}_1,\mathbf{p}_1),...,\mathcal{F}(\mathbf{t}_n,\mathbf{p}_n))$. Finally, the $n$-pairs $(\mathbf{m}_i, \mathcal{I}(\mathbf{t}_i,\mathbf{m}_i)):=(\mathcal{F}(\mathbf{t}_i,\mathbf{p}_i), \mathbf{p}_i)$, $i=1,...,n$ form our supervised database.

We denote by $\mathbf{T} \in \mathbb{R}^{n_t} \times \mathbb{R}^{n}$ to the set of all trajectory samples $(\mathbf{t}_1,..., \mathbf{t}_n)$. In other words, $\mathbf{T}$ is a matrix with $\mathbf{t}_i$ being its $i$-th column. Similarly, we define  $\mathbf{M} = (\mathbf{m}_1,..., \mathbf{m}_n) \in \mathbb{R}^{n_m} \times \mathbb{R}^{n}$ and $\mathbf{P} = (\mathbf{p}_1,..., \mathbf{p}_n) \in \mathbb{R}^{n_p} \times \mathbb{R}^{n}$.

\paragraph{Example A: Simple model problem with known analytical solution}
We select $n=10^3$ uniformly spaced samples within the subspace $[-33,33] \subset \mathbb{R}$. 

\paragraph{Example B: Inversion of borehole resistivity measurements}
 We select $n=10^6$. Then, for the five parameters described in Section~\ref{Ex:borehole}, we select random samples of the following rescaled variables over the corresponding intervals forming a subspace of $\mathbb{R}^{5}$:
	\begin{equation}
	\label{eq:7_variables_scaled}
	\begin{split}
	\log(\rho_{l}),\log(\rho_{u}),\log(\rho_{h}) &\in [0,3] \\
	\log(d_{l}), \log(d_{u}) &\in [-2,1].\\
	\end{split}
	\end{equation}

We consider arbitrary high-angle trajectories that are parameterised via the following two variables:
	\begin{equation}
	\label{eq:angle_trajectories}
	\begin{array}{lll}
	\alpha_{ini} &\in [83\degree,97\degree] &\\
	\alpha_v &\in [-0.045 \degree, 0.045\degree] &\text{(only for Example B.2)},
	\end{array}
	\end{equation}
	where $\alpha_{ini}$ is the initial trajectory dip angle and $\alpha_v$ is the variation of the dip angle at a given logging position with respect to the previous one. For each model problem, we randomly select the trajectory parameters within the above intervals. For Example B.1, $n_t=1$, while for Example B.2, $n_t=2$.

	\section{Data Preprocessing} \label{data_preprocessing}

\paragraph{Notation}
For each output parameter of $\mathcal{F}$ and $\mathcal{I}$, we denote by ${\bf x}=(x_1,...,x_n)$ the $n$-samples associated with that parameter. These $x_i$ are real scalar values for $i=1,...,n$. For example, in the borehole resistivity example, each variable ${\bf x}$ contains $n$ samples of each particular geophysical quantity such as resistivities, distances, or given measurements (attenuations, phases, etc.). Each dimension corresponds to a particular value (sample) of that variable, for example, the geosignal attenuation recorded at a specific logging position. From the algebraic point of view, the variable ${\bf x}$ denotes a row of either matrix ${\bf M}$ or ${\bf P}$.

\paragraph{Data preprocessing algorithm}
This algorithm consists of three steps. 

\begin{enumerate}
    \item {\bf Logarithmic change of coordinates}. We introduce the following change of variables:
\begin{equation}
\left.\begin{aligned}
\displaystyle {\cal R}_{\ln}({\bf x})&:=(\ln x_1, ..., \ln x_n).
\end{aligned}\right.
\label{eq:ln}
\end{equation}
For some geophysical variables (e.g., resistivity), this change of variables ensures that equal-size relative errors correspond to similar-size absolute errors. Thus, this change of variables allows us to perform {\em local} (within a variable) comparisons.

\item {\bf Remove outlier samples}. In practice, often outlier measurements are present in the sample database. These outliers appear due to measurement error or the physics of the problem. For example, in borehole resistivity measurements, some apparent resistivity measurements approach infinity, producing ``horns" in the logs. When outlier measurements exists in any particular variable of the $i$-th sample $x_i$, then the entire sample should be removed. Otherwise, outlier measurements affect the entire minimization problem, leading to poor numerical results. The removal process may be automated using statistical indicators, or decided by the user based on a priori physical knowledge about the problem. We follow this second approach in this work.

\item {\bf Linear change of coordinates}. We now introduce a linear rescaling mapping into the interval $[0.5, 1.5]$.  We select this interval since it has unit length and the mean of a normal (or a uniform) distribution variable {\bf x} is equal to one. Let $x_{\min}:=\min_{i}{x_i}$, $x_{\max}:=\max_{i}{x_i}$. We define
\begin{equation}
\label{eq:linear_scaling}
\left.\begin{aligned}
\displaystyle {\cal R}_{lin}({\bf x})&:=\left(\frac{x_1-x_{\min}}{x_{\max}-x_{\min}}+0.5, ..., \frac{x_n-x_{\min}}{x_{\max}-x_{\min}}+0.5\right),
\end{aligned}\right.
\end{equation}
where the limits $x_{\min}$ and $x_{\max}$ are fixed for all possible approximations ${\bf x}^{app}$. This change of variables allows us to perform a {\em global} comparison between errors corresponding to different variables since they all take values over the same interval.

\end{enumerate}

\paragraph{Variables classification}
We categorize each input and output geophysical variable ${\bf x}$ into two types: either linear (A) or log-linear (B). When necessary, we shall indicate that a particular variable belongs to a specific category by adding the corresponding symbol as subindex of the variable, e.g., ${\bf x}_A$. Table~\ref{tab_type_of_variables} describes the domain of those variables as well as the rescaling employed for each of them. Variables of type $A$ only require a global rescaling while those of type $B$ require both a local and a global change of variables. 
\begin{table}[!htp]
\centering
\begin{tabular}{|l||c|r|l|}
\hline
    Geophysical Variables   & Category  & Domain & Rescaling    \\
\hline
\hline
 Angles, attenuations,  & $A$ & $\mathbb{R}^n$  & ${\cal R}_{lin}({\bf x})$    \\
phases, and geosignals & &    &   \\
 \hline
Apparent resistivities, & $B$  & $(a,\infty)^n$   & ${\cal R}_{lin}({\cal R}_{\ln}({\bf x}))$   \\
resistivities, and distances &    &   $a>0$ &    \\
\hline
\end{tabular}
\caption{\label{tab_type_of_variables}
Categories for geophysical variables: types $A$ or $B$. We apply a different rescaling to each of them.}
\end{table}

For simplicity, we denote by ${\cal R}$ the result of the above rescalings, i.e., ${\cal R}({\bf x}_A):={\cal R}_{lin}({\bf x}_A)$, and ${\cal R}({\bf x}_B):={\cal R}_{lin}({\cal R}_{\ln}({\bf x}_B))$. In general, given a variable ${\bf x}$ (of category $A$ or $B$), we represent ${\bf x}_{\cal R}:={\cal R}({\bf x})$. Given a matrix ${\bf X} \in \mathbb{R}^{n_x} \times \mathbb{R}^{n}$, we abuse notation and denote by ${\bf X}_{\cal R}:={\cal R}({\bf X}) \in \mathbb{R}^{n_x} \times \mathbb{R}^{n}$ to the matrix that results from applying operator ${\cal R}$ rowwise.

{\bf Remark:}  Substituting in Equation~\ref{eq:ln} the Napierian logarithm for the base ten logarithm does not affect the definition of ${\cal R}$. Results are identical.

	\section{Norms and Errors} \label{norms_and_errors}

We first introduce both the vector and the matrix norms that we use during the training process.

\paragraph{Vector norms}
We introduce a norm $|| \cdot ||_{\mathbb{X}}$ associated with the variable ${\bf x}$. In general, we employ the $l_1$ or $l_2$ norms given by:
\begin{equation}
\label{eq:norms}
\left.\begin{aligned}
|| {\bf x} ||_1 &= \sum_{i=1}^n  | x_i|,\\
|| {\bf x} ||_2 &= \sqrt{\sum_{i=1}^n  | x_i|^2}.
\end{aligned}\right.
\end{equation}

\paragraph{Matrix norms}
We introduce a norm $|| \cdot ||_X$ associated with matrix ${\bf X}=(x_{ij}) \in \mathbb{R}^{n_x} \times \mathbb{R}^{n}$. We consider an ``entrywise norm" of the type $l_{p,q}$ for some $p,q \geq 1$ defined for a matrix ${\bf X}$ as (see~\cite{rahimpour2017feature}):
	\begin{equation}
	\label{eq:matrix_norm}
	|| \mathbf{X} ||_{p,q} := \left( \sum_{i=1}^{n_x}  \left( \sum_{j=1}^n | x_{ij} |^q \right)^{\frac{p}{q}} \right)^{\frac{1}{p}} .
	\end{equation}
	In this work, we employ the $l_{1,1}$ and $l_{2,2}$ norms, where the $l_{2,2}$ norm is the Frobenius norm. 

\paragraph{Absolute and relative errors}
Let ${\bf x}^{app}=(x_1^{app},...,x_n^{app})$ be an approximation of ${\bf x}$. We define the absolute error $A_e$ between ${\bf x}^{app}$ and ${\bf x}$ in the $|| \cdot ||_{\mathbb{X}}$ norm as
\begin{equation}
\left.\begin{aligned}
A_e^{\mathbb{X}}({\bf x}^{app},{\bf x}) &:=    || {\bf x}^{app} - {\bf x} ||_{\mathbb{X}}.
\end{aligned}\right.
\end{equation}
Similarly, for a matrix ${\bf X}$ and its approximation ${\bf X}^{app}$, we define:
\begin{equation}
\left.\begin{aligned}
A_e^{X}({\bf X}^{app},{\bf X}) &:=    || {\bf X}^{app} - {\bf X} ||_{X}.
\end{aligned}\right.
\end{equation}

This error measure has limited use since it is challenging to select an absolute error threshold that distinguishes between a {\em good} and a {\em bad} quality approximation. To overcome this issue, practitioners often employ relative errors. We define the relative error $R_e$ in percent between ${\bf x}^{app}$ and ${\bf x}$ in the $|| \cdot ||_{\mathbb{X}}$ norm as:
\begin{equation}
\left.\begin{aligned}
R_e^{\mathbb{X}}({\bf x}^{app},{\bf x}) &:=  100  \frac{|| {\bf x}^{app} - {\bf x} ||_{\mathbb{X}}}{|| {\bf x} ||_{\mathbb{X}}}.
\end{aligned}\right.
\end{equation}
Similarly, for matrices we define:
\begin{equation}
\left.\begin{aligned}
R_e^X({\bf X}^{app},{\bf X}) &:=  100  \frac{|| {\bf X}^{app} - {\bf X} ||_X}{|| {\bf X} ||_X}.
\end{aligned}\right.
\end{equation}

\paragraph{Error control}
For a variable ${\bf x}$ and its approximation ${\bf x}^{app}$, we want to control the relative error of the rescaled variable, that is:
\begin{equation}
\left.\begin{aligned}
R_e^{\mathbb{X}}({\bf x}^{app}_{\cal R},{\bf x}_{\cal R}).
\end{aligned}\right.
\end{equation}

The value $|| {\bf x}_{\cal R} ||_{\mathbb{X}}$ is expected to be similar for all variables ${\bf x}$. Thus, the minimizer of the sum over the existing variables of the absolute errors $A_e^{\mathbb{X}}({\bf x}^{app}_{\cal R},{\bf x}_{\cal R})$ delivers almost optimal results in terms of minimizing the sum of relative errors.

	\section{Loss Function} \label{loss_function}
In this section, we consider a set of weights $\theta\in\Theta$ and a function $\mathcal{I}_{{\cal R},\theta}$ that depend upon the selected DNN architecture (see Appendix~\ref{sec:dnn_arquitectures}). Then, we introduce a loss function that depends on $\mathcal{I}_{{\cal R},\theta}$. We denote as $\mathcal{I}_{{\cal R},\theta^\ast}$ to the minimizer of the loss function over all possible weight sets. Function $\mathcal{I}_{\theta^\ast}:={\cal R}^{-1} \circ \mathcal{I}_{{\cal R},\theta^\ast} \circ {\cal R}$ is the final DNN approximation of $\mathcal{I}$. By repeating this process with different loss functions, we analyze the advantages and limitations of each one.

\subsection{Data misfit}
	A simple loss function (and the corresponding optimization problem) based on the data misfit is given by:
	\begin{equation}
	\label{eq:loss1}
	\mathcal{I}_{{\cal R},\theta^\ast}:=\arg \min_{\theta\in\Theta} || \mathcal{I}_{{\cal R},\theta}(\mathbf{T}_{\cal R},\mathbf{M}_{\cal R})- \mathbf{P}_{\cal R}||_{P}.
	\end{equation}
	In the above equation, symbol $|| \cdot ||_{P}$ indicates a matrix norm of the type introduced in Section~\ref{norms_and_errors}. 
	
\subsubsection{Example A: Model problem with known analytical solution}
\begin{figure}[!h]
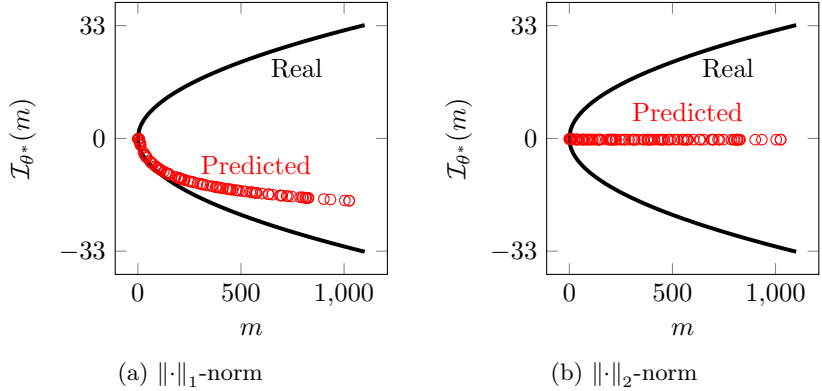

\centering
	\subcaptionbox{$\left\Vert \cdot \right\Vert_{1}$-norm \label{fig:loss1_norm2}}{%
	\input{x2_n/Fig_loss1_norm1.tex}}
	\hspace{0.3cm}
	\subcaptionbox{$\left\Vert \cdot \right\Vert_{2}$-norm \label{fig:loss1_norm1}}{%
	\input{x2_n/Fig_loss1_norm2.tex}}
	\caption{Analytical solution vs DNN predicted solution evaluated over the test dataset using the loss function based on the data misfit.}
	\label{fig:loss1}
\end{figure}

In this example, $n_p=1$. Thus, matrix norms reduce to vector norms. Figure~\ref{fig:loss1} illustrates the results we obtain using the $l_1$ and $l_2$ norms, respectively. These disappointing results are expected. Indeed, for the $l_2$-norm, it is easy to show that for a sufficiently flexible DNN architecture, the exact solution is $\mathcal{I}_{\theta^\ast} \approx 0$. To prove this, we assume that for every sample of the form $(m, \sqrt{m})$, there exist another one $(m, -\sqrt{m})$, which is satisfied in our dataset by construction (see Section~\ref{data_space_and_GT}). Then, for each pair of samples of this form, the exact point that minimizes the distance between both solutions ($\sqrt{m}$ and $-\sqrt{m}$) is 0. This argument can be extended to all pairs of samples. A similar reasoning shows that for the $l_1$-norm, any solution in between the two square root branches is an exact solution of the inverse problem. Our numerical solutions in Figure~\ref{fig:loss1} confirm these simple mathematical observations. Thus, the data misfit loss function is unsuitable for inversion purposes.
	
\subsection{Misfit of the measurements}
	To overcome the aforementioned limitation, we consider the following loss function that measures the misfit of the measurements (see~\cite{jin2019using}):
	\begin{equation}
	\label{eq:loss2}
	\mathcal{I}_{{\cal R},\theta^\ast}:=\arg \min_{\theta\in\Theta}\|(\mathcal{F}_{\cal R} \circ \mathcal{I}_{{\cal R},\theta})(\mathbf{T}_{\cal R},\mathbf{M}_{\cal R})-\mathbf{M}_{\cal R}\|_M,
	\end{equation}
	where $\mathcal{F}_{\cal R} := {\cal R} \circ \mathcal{F} \circ {\cal R}^{-1}$, and $|| \cdot ||_{M}$ indicates a matrix norm of the type introduced in Section~\ref{norms_and_errors}.
	
\begin{figure}[!h]
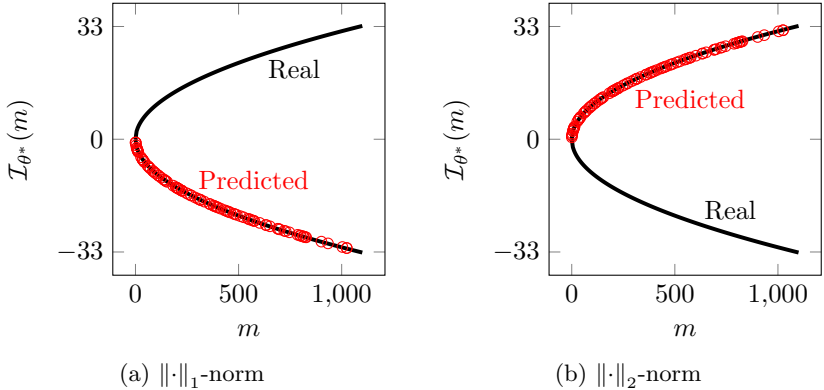

\centering
	\subcaptionbox{$\left\Vert \cdot \right\Vert_{1}$-norm}{%
	\input{x2_n/Fig_loss2_norm1.tex}} 
	\hspace{0.3cm}
	\subcaptionbox{$\left\Vert \cdot \right\Vert_{2}$-norm}{%
	\input{x2_n/Fig_loss2_norm2.tex}}
	\caption{Analytical solution vs DNN predicted solution evaluated over the test dataset using the loss function based on the measurements misfit.}
	\label{fig:loss2}
\end{figure}

 \paragraph{Example A: Model problem with known analytical solution}
Figure~\ref{fig:loss2} shows the inversion results when using the misfit of the measurements. We recover one of the possible solutions of the inverse operator. A regularization term could be introduced to select one solution branch over the other.

Despite the accurate results exhibited for the above example, the proposed loss function has some critical limitations that affect its performance. Namely, during training, it is necessary to evaluate the forward problem multiple times. Depending upon the size of the training dataset and number of iterations required to converge, this may lead to millions of forward function evaluations. Solving the forward problem for such large number of times is time-consuming even with a 1.5D semi-analytic simulator. Moreover, most forward solvers are implemented for CPU architectures, while the training of the DNN normally occurs on GPUs. This requires a permanent communication between GPU and CPU, which further slows down the training process. Additionally, porting the forward solver $\mathcal{F}$ to a GPU may be complex to implement and bring additional numerical difficulties.

\subsection{Encoder-Decoder}

	To overcome the aforementioned implementation challenges, we propose to approximate the forward function using another DNN $\mathcal{F}_{\phi^\ast}$, where $\phi^\ast \in \Phi$ are the parameters associated to the trained DNN. With this approach, we simultaneously train the forward and inverse operators solving the following optimization problem:
\begin{equation}
\begin{aligned}
	(\mathcal{F}_{{\cal R},\phi^\ast}, \mathcal{I}_{{\cal R},\theta^\ast}):=\arg \min_{\phi \in \Phi, \theta \in \Theta} \{ & \|(\mathcal{F}_{{\cal R},\phi} \circ \mathcal{I}_{{\cal R},\theta})(\mathbf{T}_{\cal R},\mathbf{M}_{\cal R})-\mathbf{M}_{\cal R}\|_M \\
	& +\|{\cal F}_{{\cal R},\phi}(\mathbf{T}_{\cal R},{\bf P}_{\cal R})-{\bf M}_{\cal R}\|_M \},
\end{aligned}
\label{eq:loss3}
\end{equation}		
Function $\mathcal{F}_{\phi^\ast}:={\cal R}^{-1} \circ \mathcal{F}_{{\cal R},\phi^\ast} \circ {\cal R}$ is the final DNN approximation to $\mathcal{F}$. The first term in the above loss function constitutes an Encoder-Decoder DNN architecture~\cite{Badrinarayanan} and ensures that function $\mathcal{I}_{{\cal R},\theta^\ast}$ shall be a inverse of $\mathcal{F}_{{\cal R},\phi^\ast}$. The second term imposes that the forward DNN approximates the ground truth data. In particular, it prevents situations in which both $\mathcal{I}_{{\cal R},\theta^\ast}$ and $\mathcal{F}_{{\cal R},\phi^\ast}$ approximate the identity operator.

\begin{figure}[!h]
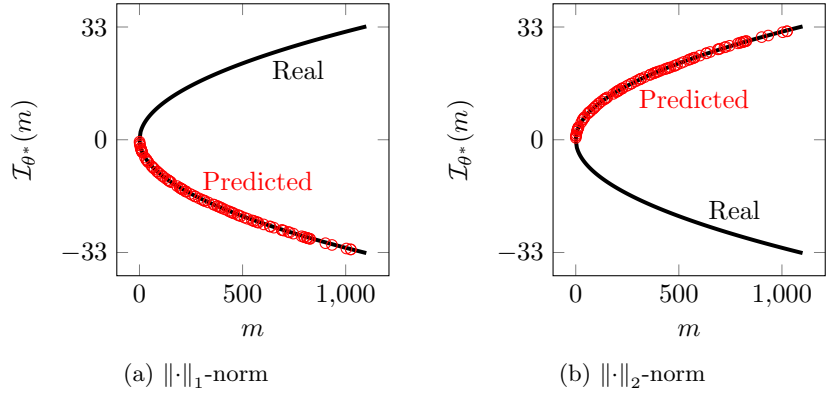

	\centering
	\subcaptionbox{$\left\Vert \cdot \right\Vert_{1}$-norm}{%
		\input{x2_n/Fig_loss4_norm1_INV.tex}} 
	\hspace{0.3cm}
	\subcaptionbox{$\left\Vert \cdot \right\Vert_{2}$-norm}{%
		\input{x2_n/Fig_loss4_norm2_INV.tex}}
	\caption{Analytical solution vs DNN predicted solution evaluated over the test dataset using the Encoder-Decoder loss function.}
	\label{fig:loss3}
\end{figure}

 \paragraph{Example A: Model problem with known analytical solution}
Figure~\ref{fig:loss3} shows the results obtained with the Encoder-Decoder loss function. We recover accurate inversion results.

\subsection{Two-steps approach}
It is possible to decompose the above Encoder-Decoder based loss function into two step: the first one intended to approximate the forward function, and the second one to determine the inverse operator:
\begin{equation}
\label{eq:loss4}
\begin{aligned}
	\mathcal{F}_{{\cal R},\phi^\ast}:= & \arg \min_{\phi \in \Phi} \|{\cal F}_{{\cal R},\phi}(\mathbf{T}_{\cal R},{\bf P}_{\cal R})-{\bf M}_{\cal R}\|_M,\\
	 \mathcal{I}_{{\cal R},\theta^\ast}:= & \arg \min_{\theta \in \Theta}\|(\mathcal{F}_{{\cal R},\phi^\ast} \circ \mathcal{I}_{{\cal R},\theta})(\mathbf{T}_{\cal R},\mathbf{M}_{\cal R})-\mathbf{M}_{\cal R}\|_M.
\end{aligned}
\end{equation}	
	
\begin{figure}[!h]
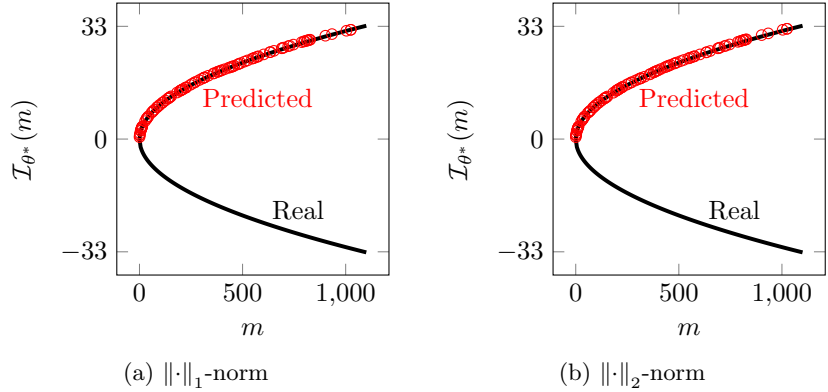

	\centering
	\subcaptionbox{$\left\Vert \cdot \right\Vert_{1}$-norm}{%
		\input{x2_n/Fig_loss5_norm1_INV.tex}} 
	\hspace{0.3cm}
	\subcaptionbox{$\left\Vert \cdot \right\Vert_{2}$-norm}{%
		\input{x2_n/Fig_loss5_norm2_INV.tex}}
	\caption{Analytical solution vs DNN predicted solution evaluated over the test dataset using the two-step loss function.}
	\label{fig:loss4}
\end{figure}

 \paragraph{Example A: Model problem with known analytical solution}
Figure~\ref{fig:loss4} shows the results of the inversion using the two-steps approach. We recover a faithful approximation of the inverse operator. 

{\bf Remark A:} Based on the above discussion, it may seem that loss functions given by either Equations~\ref{eq:loss3} or~\ref{eq:loss4} are ideal to solve inverse problems. However, there is a critical issue that needs to be addressed. In Equation~\ref{eq:loss4}a, the forward DNN  ${\cal F}_{{\cal R},\phi}$ is trained only for the given dataset samples. However, the output of the DNN approximation of the inverse operator $\mathcal{I}_{{\cal R},\theta}$ will often deliver data far away from the data space used to produce the training samples. This may lead to catastrophic results. To illustrate this, we consider our model problem with known analytical solution. If we consider a dataset with only positive values of $p$, then the following approximations will lead to a zero loss function:
\begin{equation}
\mathcal{F}_{\phi^\ast}(p)= \left\{
\begin{array}{ll}
  p^2 & \textrm{if $p>0$} \\
  a p^2 & \textrm{if $p<0$}  
\end{array}
\right.
\hspace{0.2in}
\mathcal{I}_{\theta^\ast}(m)=   -\sqrt{m/a},
\label{exampleA_bad_result}
\end{equation}
for any $a>0$. However, if $a \neq 1$, this approximation is far away from any of our two original solutions of the inverse problem. To prevent these undesired situations, one should ensure that the output space of $\mathcal{F}_{{\cal R},\theta^\ast}$  is sufficiently close to the space from which we obtain the training samples. However, this is often difficult to control.

\subsection{Regularization term}
Inverse problems often exhibit non-unique solutions. Thus, in numerical methods, one introduces a regularization term to select a particular solution we prefer out of all the existing ones. 

In DL applications, standard regularization techniques seek to optimize the model architecture (e.g., by penalizing high-valued weights). Herein, we regularize the system by adding the term given by the loss function of Equation~\ref{eq:loss1} measured in the $l_1$-norm to either the loss function given by Equation~\ref{eq:loss3} or~\ref{eq:loss4}b. This extra term guides the solution towards the ones considered in the training dataset, which may be convenient. Nevertheless, such a regularization term often hides the fact that other different solutions of the inverse problem may coexist. We study the advantages and limitations of including this regularization term in detail in Section~\ref{sec:numerical_results}.

	\section{Implementation \label{implementation}}

To solve the forward problem, we employ a semi-analytic method~\cite{Loseth} implemented in Fortran 90~\cite{Shahriari}. With it, we produce a dataset containing one million samples ({\em ground truth}). Each sample consists of a randomly selected 1D layered model (see Section~\ref{data_space_and_GT} for details). We use 80\% of the samples for training the DNNs, 10\% for validating them, and the remaining 10\% for testing. 

We consider two DNN architectures to approximate $\mathcal{F}$ and $\mathcal{I}$, respectively. The forward function $\mathcal{F}$ is well-posed and continuous, while the inverse operator $\mathcal{I}$ is not even well-defined. Thus, we employ a simpler DNN architecture to approximate $\mathcal{F}$ than to approximate $\mathcal{I}$. See Appendix~\ref{sec:dnn_arquitectures} for details.

We implement our DNNs using \textit{Tensorflow 2.0}~\cite{tf2} and \textit{Keras}~\cite{chollet2015}  libraries. To train the DNNs, we use a \textit{NVIDIA Quadro GV100} GPU. Using this hardware device, we require almost 70 hours to simultaneously train $\mathcal{F}_{{\cal R},\phi^\ast}$ and $\mathcal{I}_{{\cal R},\theta^\ast}$. While the training process is time-consuming, it is performed \textit{offline}. Then, the \textit{online} part of the process consists of simply evaluating the DNN, which can deliver an inverse model for thousands of logging positions in a few seconds. This low \textit{online} computational cost makes the DNN approach an excellent candidate to perform inversion during geosteering operations in the field.

	\section{\bf Numerical Results \label{sec:numerical_results}}

We perform a three step evaluation process of the results:
\begin{enumerate}
\item We first study the evolution of each term in the loss function during the training process. This analysis assesses the overall performance of the training process and, in particular, shows if any particular term of the loss function is driving the optimization procedure in detriment of other terms.
\item Second, we produce multiple cross-plots, which provide essential information about the adequacy of the selected loss function and dataset. These cross-plots indicate the possible non-uniqueness of the inverse problem at hand.
\item Finally, we apply the trained networks to invert three realistic synthetic models and analyze the overall success of the proposed DNN algorithm as well as its limitations.
\end{enumerate}
The above evaluation process provides a step-by-step assessment of the adequacy of the proposed strategy for solving inverse problems.

In most cases, we observe similar results when we consider the Encoder-Decoder loss function given by Equation~\ref{eq:loss3} and the two-step loss function given by Equation~\ref{eq:loss4}. For brevity, we mostly focus on the Encoder-Decoder results. Additionally, we include one set of results using the two-step loss function, for which the observed behavior is essentially different from that of the Encoder-Decoder process.

\subsection{Evolution of the loss function}

\begin{figure}[!h]
	\centering
	{%
		\subcaptionbox{$\|{\cal F}_{{\cal R},\phi}(\mathbf{T}_{\cal R},{\bf P}_{\cal R})-{\bf M}_{\cal R}\|_M$}{\includegraphics[width=0.45\textwidth]{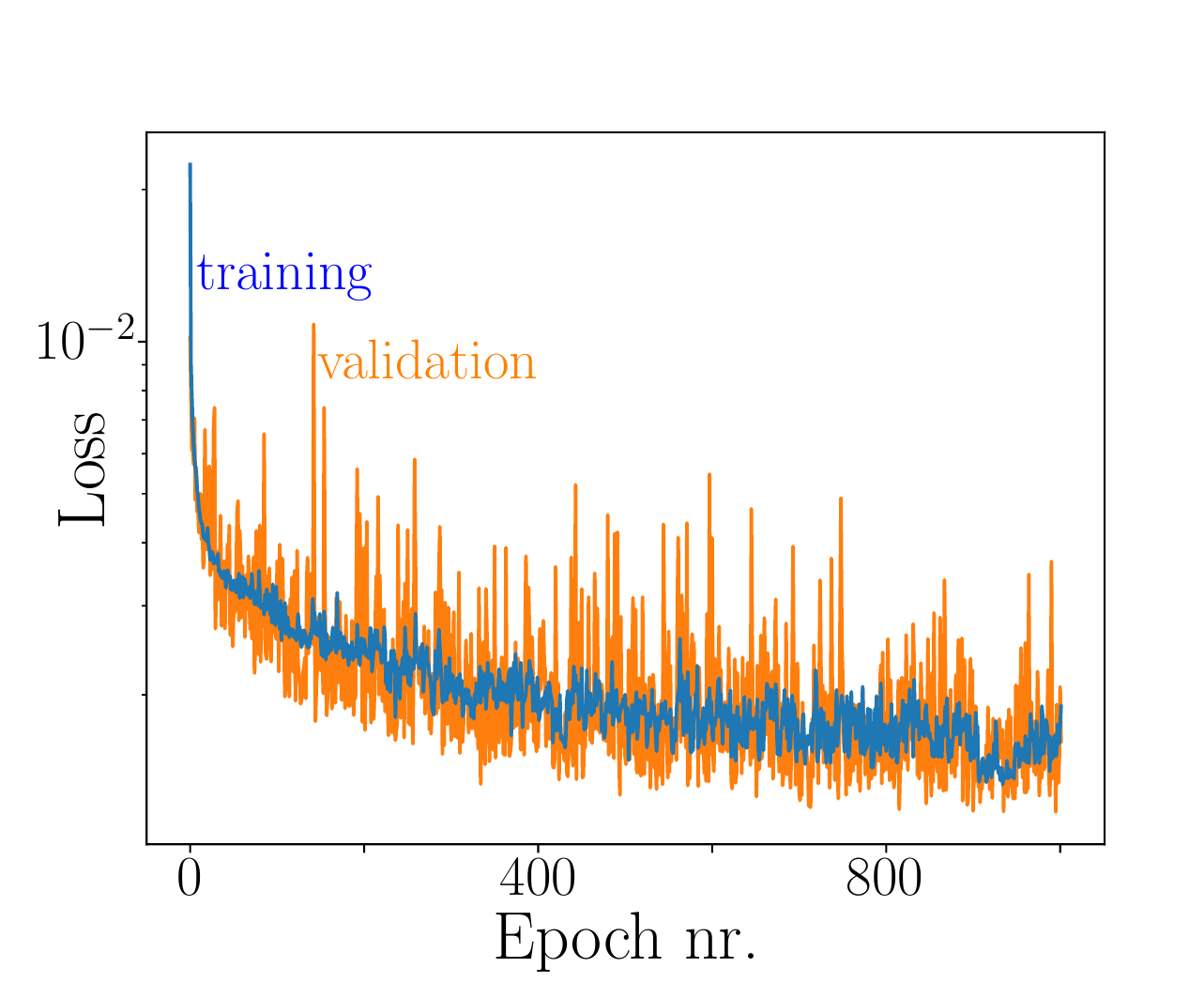}}
		\hspace{0.1cm}
		\subcaptionbox{$\|(\mathcal{F}_{{\cal R},\phi} \circ \mathcal{I}_{{\cal R},\theta})(\mathbf{T}_{\cal R},\mathbf{M}_{\cal R})-\mathbf{M}_{\cal R}\|_M$}{\includegraphics[width=0.46\textwidth]{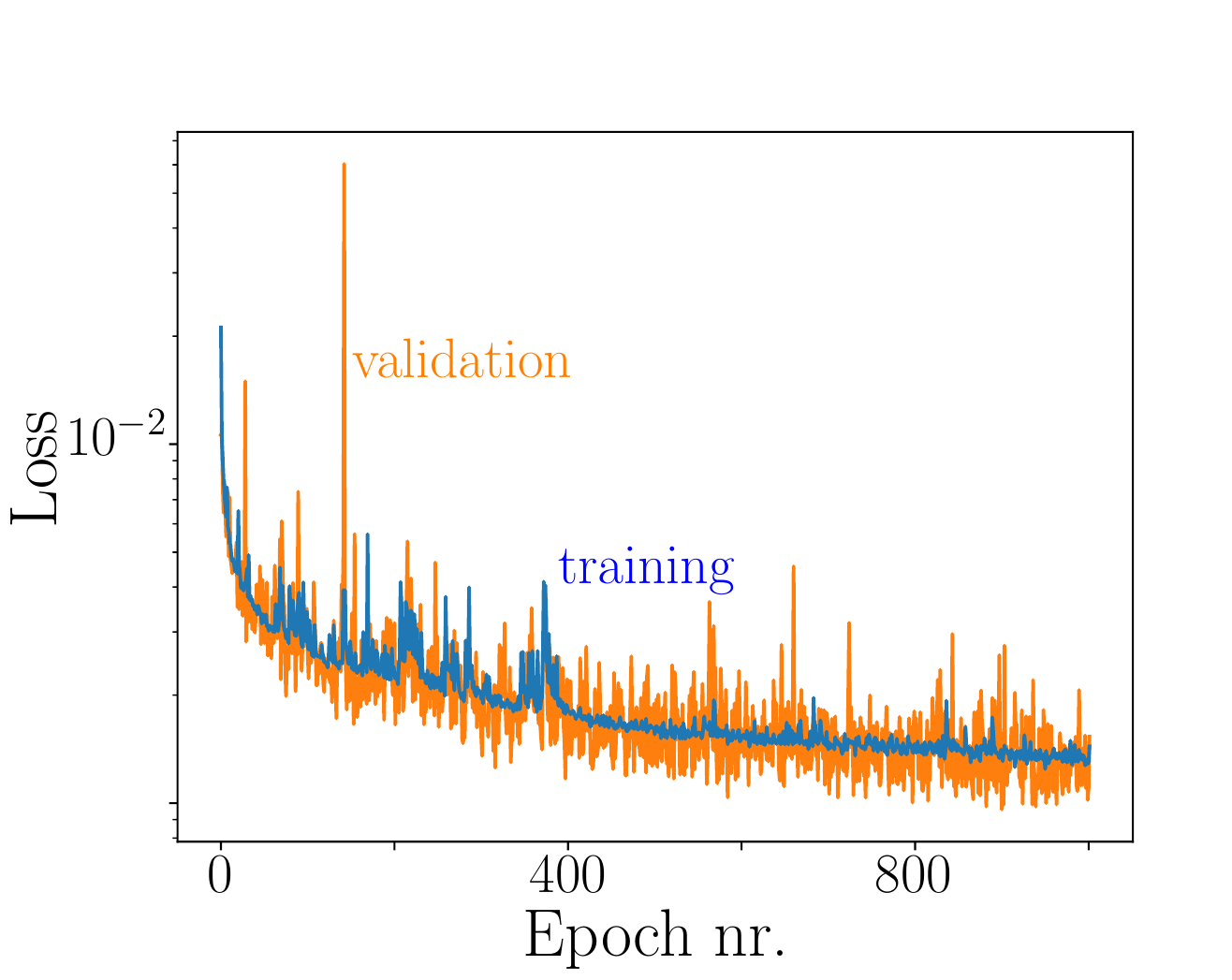}}}	
	{%
		\subcaptionbox{Total Loss}{\includegraphics[width=0.45\textwidth]{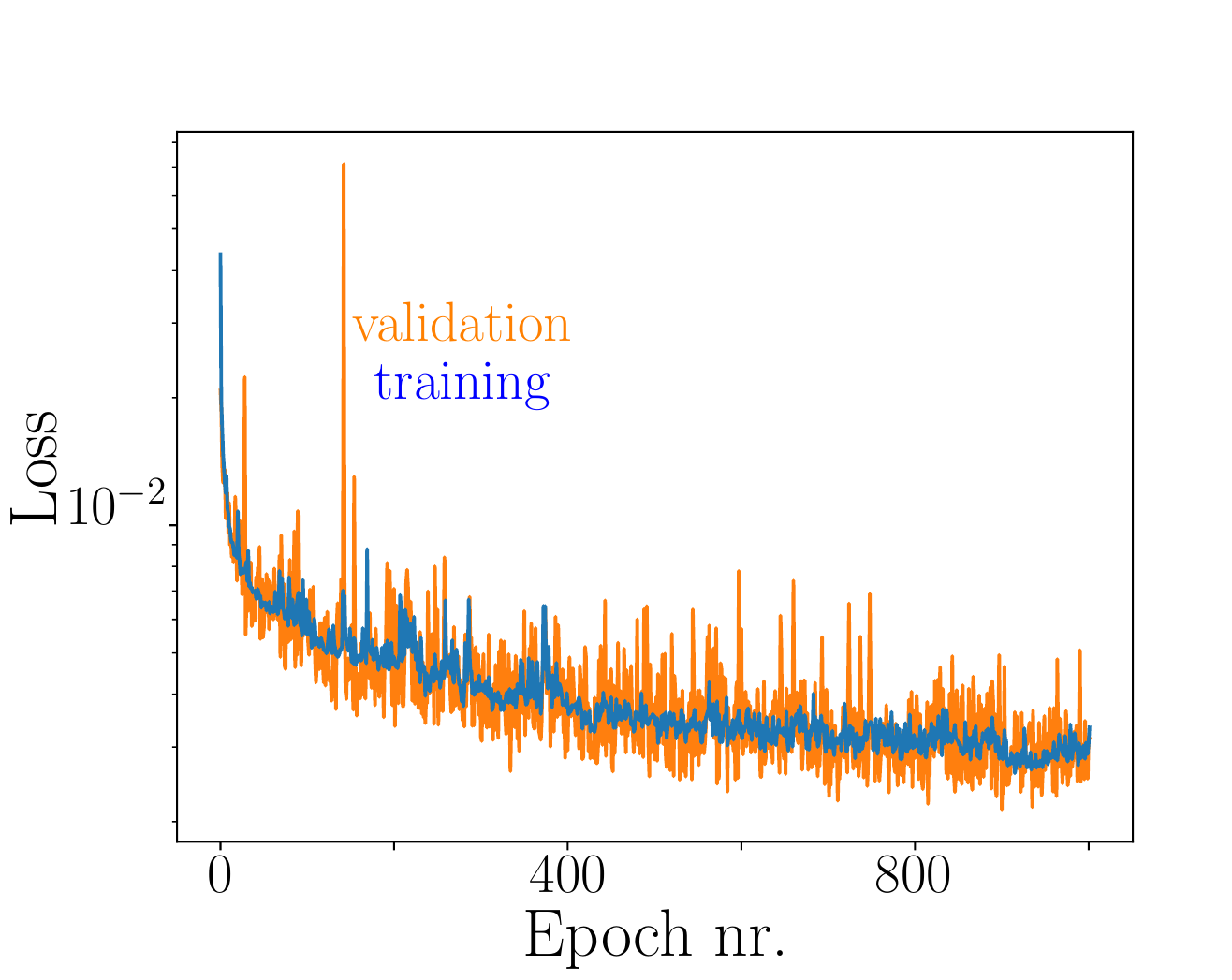}}}
	\caption{\label{fig:loss_evolution_without_regularization}Example B.1. Evolution of the different terms of the Encoder-Decoder loss function given by Equation~\ref{eq:loss3} without regularization.}
\end{figure}

\begin{figure}[!h]
	\centering
	{%
		\subcaptionbox{$\|{\cal F}_{{\cal R},\phi}(\mathbf{T}_{\cal R},{\bf P}_{\cal R})-{\bf M}_{\cal R}\|_M$}{\includegraphics[width=0.45\textwidth]{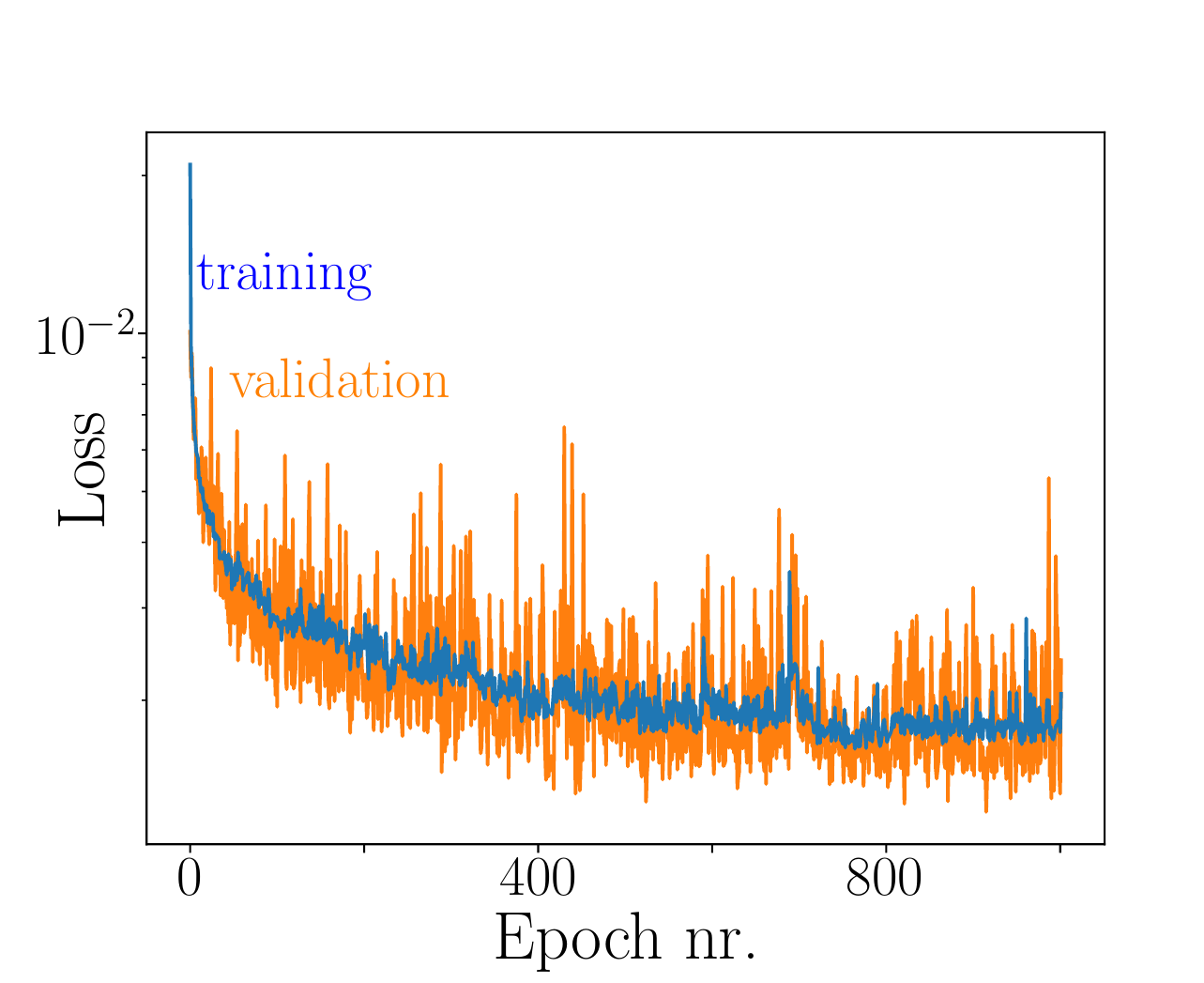}}
		\hspace{0.1cm}
		\subcaptionbox{$\|(\mathcal{F}_{{\cal R},\phi} \circ \mathcal{I}_{{\cal R},\theta})(\mathbf{T}_{\cal R},\mathbf{M}_{\cal R})-\mathbf{M}_{\cal R}\|_M$}{\includegraphics[width=0.46\textwidth]{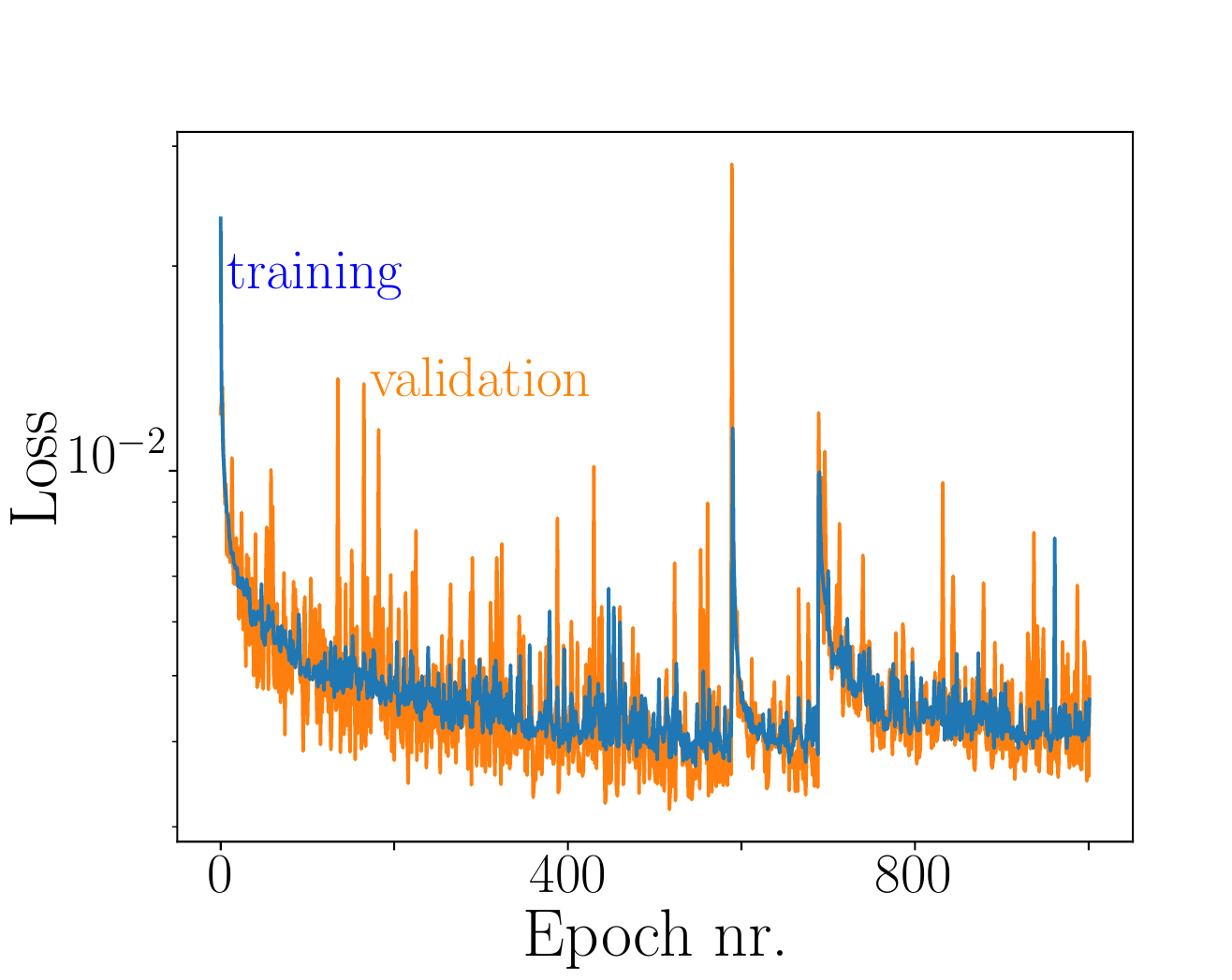}}
}
	{%
		\subcaptionbox{$ || \mathcal{I}_{{\cal R},\theta}(\mathbf{T}_{\cal R},\mathbf{M}_{\cal R})- \mathbf{P}_{\cal R}||_{P}$}{\includegraphics[width=0.45\textwidth]{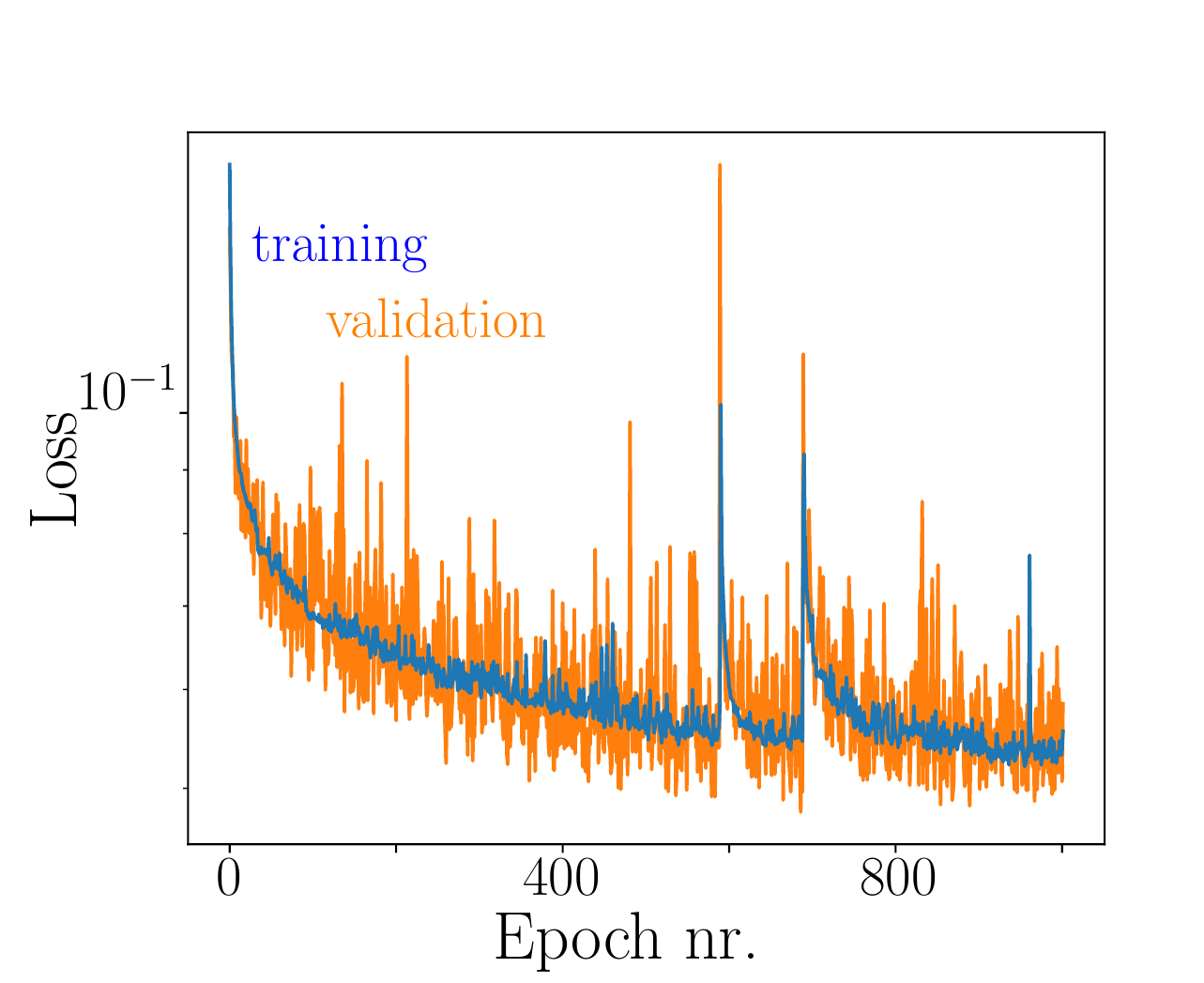}}
		\hspace{0.1cm}
		\subcaptionbox{Total Loss}{\includegraphics[width=0.45\textwidth]{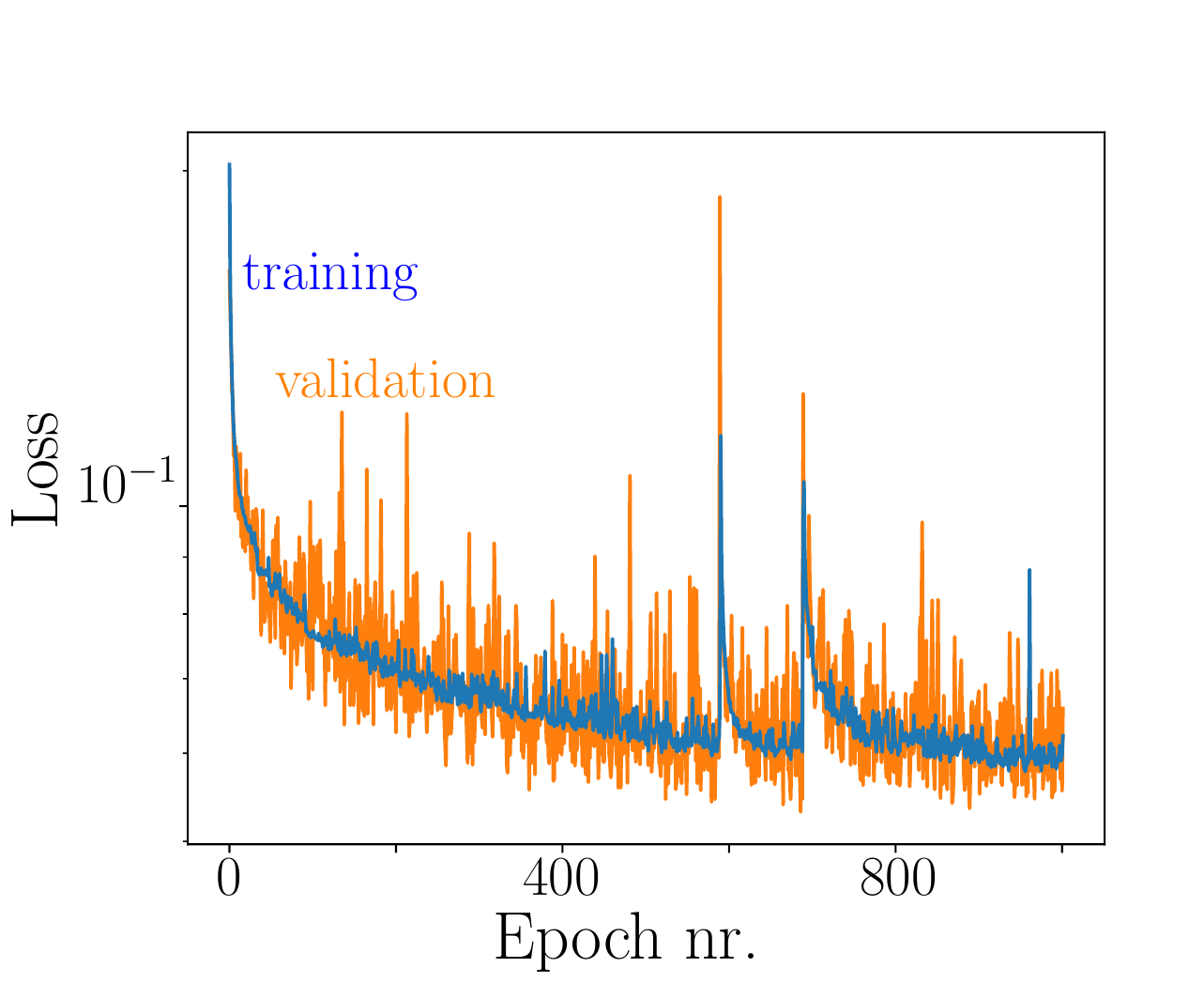}}}
	\caption{\label{fig:loss_evolution_with_regularization}Example B.1. Evolution of the different terms of the Encoder-Decoder loss function given by Equation~\ref{eq:loss3} with the regularization term prescribed by Equation~\ref{eq:loss1}.}
\end{figure}

Figure~\ref{fig:loss_evolution_without_regularization} displays the evolution of the terms composing the Encoder-Decoder loss function described in Equation~\ref{eq:loss3} for Example B.1. Figure~\ref{fig:loss_evolution_with_regularization} displays the corresponding results when we add the regularization term based on Equation~\ref{eq:loss1}. In both figures, we observe: (a) a proper reduction of the total loss function, indicating that the overall minimization process is successful; (b) an adequate balance between the loss contribution of the different terms composing each loss function, suggesting that all terms of the loss functions are simultaneously minimized; and (c) a satisfactory match between the loss functions corresponding to the training and the validation data samples, which indicates we avoid overfitting. We observe a similar behavior with Example B.2, which we skip for brevity. We do not detail the results per variable since the applied rescaling of Section~\ref{data_preprocessing} guarantees a good balance between different variables.

\subsection{Cross-plots}

We consider the following types of cross-plots:
\begin{equation}
\begin{array}{lrcr}
\text{Cross-plot 1:} &  \mathcal{F} \circ \mathcal{I} &  vs &  \mathcal{F}_{\phi^\ast} \circ \mathcal{I} \\ 
 \text{Cross-plot 2:} &  \mathcal{F} \circ \mathcal{I} &  vs &  \mathcal{F}_{\phi^\ast} \circ \mathcal{I}_{\theta^\ast} \\
\text{Cross-plot 3:} &   \mathcal{F} \circ \mathcal{I} &  vs &  \mathcal{F} \circ \mathcal{I}_{\theta^\ast} \\
\text{Cross-plot 4:} &  \mathcal{I} &  vs & \mathcal{I}_{\theta^\ast}
 \end{array}
 \label{eq:crossplots}
\end{equation}
In the above, $\mathcal{F}$ and $\mathcal{I}$ are the exact functions and they define the {\em ground truth}, while the others are the {\em predictions} our DNNs deliver. In particular, in the first three types of cross-plots the ground truth is simply the identity mapping. We could display each type of cross-plot for the training, validation, and test data samples and for each variable. In our Example B, this makes a total of 69 cross-plots. In addition, we need to repeat them for each considered loss function. To compress this information, we quantify each cross-plot with a single number: the statistical measure $R$-squared ($R^2$), which represents how much variation of the ground truth is explained by the predicted value. When this value is close to $1$, indicating a perfect matching between the predicted value and the ground truth, we can safely omit these cross-plots. Otherwise, cross-plots display interesting information beyond what $R^2$ provides. 

The proper interpretation of the cross-plots (or alternatively, $R^2$ factors) is of utmost importance. Cross-plots of type 1 (Equation~\ref{eq:crossplots}$_1$) indicate how well the forward function is approximated over the given dataset. The cross-plots of type 2 (Equation~\ref{eq:crossplots}$_2$) display how well the composition of the predicted forward and inverse mappings approximate the identity. These two types of cross-plots often deliver high $R^2$ factors, since the corresponding approximations are directly built into the Encoder-Decoder loss function given by Equation~\ref{eq:loss3}. Table~\ref{tab:cp12_b1} confirms those theoretical predictions for the most part.
\begin{table}[!h]
Cross-plots 1 \\
\begin{tabular}{ |l|c|c|c|c|c|c| } 
 \hline
                                             & Atten.          &  Atten. & Atten. & Phase & Phase  & Phase \\ 
     $R^2$ factors                  & LWD              &  Deep & Deep & LWD & Deep & Deep \\ 
                                             & Coaxial         & Coaxial &  Geosignal & Coaxial & Coaxial & Geosignal \\ 
 \hline
 Example B.1 & & & & & & \\
 Training         &  0.9997           &  0.9992       & {\bf 0.9509}                & 0.9996              & 0.9994         & {\bf 0.9468} \\ 
 Test            &  0.9995           &  0.9984       & {\bf 0.9531}                & 0.9990              & 0.9991         & {\bf 0.9487} \\ 
 Without Reg. & & & & & & \\ 
 \hline
 Example B.1 & & & & & & \\
  Training         &  0.9998           &  0.9998       & 0.9897                & 0.9998              & 0.9998         & 0.9893 \\ 
 Test            &  0.9998           &  0.9998       & 0.9893                & 0.9998              & 0.9998         & 0.9890 \\ 
 With Reg. & & & & & & \\ 
 \hline
  Example B.2 & & & & & & \\
  Training         &  0.9959           &  0.9975       & 0.9872                & 0.9954              & 0.9980        & 0.9853 \\ 
  Test            &  0.9924           &  0.9960       & 0.9775                & 0.9920              & 0.9974         & 0.9765 \\ 
 Without Reg. & & & & & & \\ 
\hline
\end{tabular}
\\
\\
Cross-plots 2\\
\begin{tabular}{ |l|c|c|c|c|c|c| } 
\hline
                                             & Atten.          &  Atten. & Atten. & Phase & Phase  & Phase \\ 
     $R^2$ factors                  & LWD              &  Deep & Deep & LWD & Deep & Deep \\ 
                                             & Coaxial         & Coaxial &  Geosignal & Coaxial & Coaxial & Geosignal \\ 
 \hline
  Example B.1 & & & & & & \\
 Training       &  0.9997           &  0.9995       & 0.9998                & 0.9999              & 0.9996         & 0.9999 \\ 
 Test           &  0.9997           &  0.9994       & 0.9999                & 0.9999            & 0.9996         & 0.9999 \\ 
 Without Reg. & & & & & & \\ 
 \hline
 Example B.1 & & & & & & \\
 Training       &  0.9971           &  0.9980       & 0.9779                & 0.9970              & 0.9979         & 0.9798 \\ 
Test           &  0.9970           &  0.9979       & 0.9785                & 0.9970              & 0.9978         & 0.9803 \\ 
 With Reg. & & & & & & \\ 
 \hline
   Example B.2 & & & & & & \\
 Training         &  0.9931           &  0.9958       & 0.9800                & 0.9933              & 0.9967        & 0.9821 \\ 
 Test            &  0.9890           &  0.9930       & 0.9701                & 0.9881              & 0.9944         & 0.9720 \\ 
 Without Reg. & & & & & & \\ 
 \hline
\end{tabular}
\caption{\label{tab:cp12_b1}$R^2$ factors for cross-plots 1 and 2 and Examples B.1 and B.2, with and without regularization, for training and test datasets. Numbers below 0.96 are marked in boldface.}
\end{table}

\begin{figure}[!h]
\centering
	{%
	\includegraphics[scale=0.9]{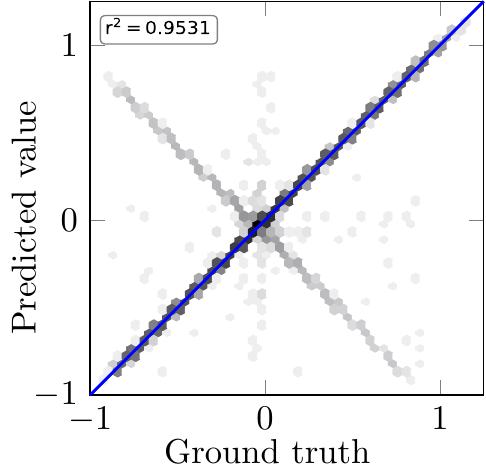}
	\hspace{0.2cm}
	\includegraphics[scale=0.9]{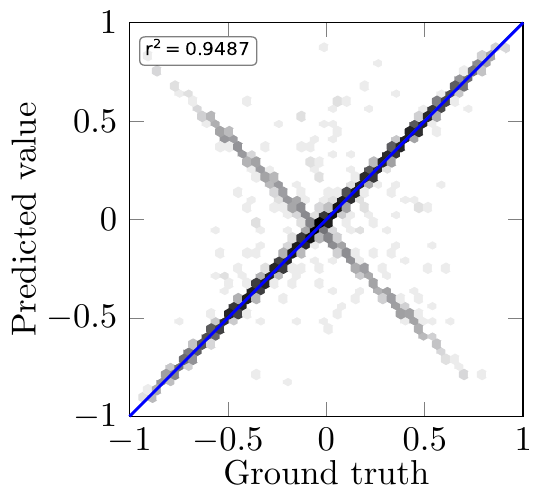}}
	{%
	\includegraphics[scale=0.9]{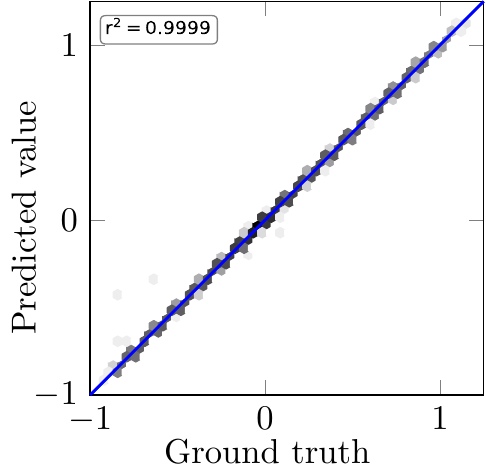}
	\hspace{0.2cm}
	\includegraphics[scale=0.9]{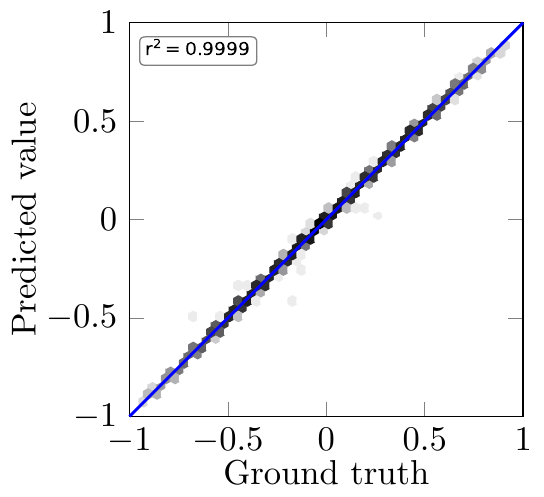}}
	\caption{\label{fig:cross-plots1-2_geosignal} Geosignal cross plots for the Example B.1 without regularization for the test dataset. First row: Cross-plots 1. Second row: Cross-plots 2. First column: Attenuation. Second column: Phase.}
\end{figure}

An in-depth inspection of Table~\ref{tab:cp12_b1} reveals that for the the geosignal measurements (both attenuation and phase) corresponding to the Example B.1 without regularization, the cross-plots 2 exhibit significantly better $R^2$ factors than those corresponding to the cross-plots 1. Figure~\ref{fig:cross-plots1-2_geosignal} shows the corresponding cross-plots. The anti-diagonal grey line shown in cross-plots of type 1 corresponds to dip angles of the logging instrument that are close to 90 degrees. At that angle, the geosignal is discontinuous (see Appendix~\ref{app:meas}). Thus, it is not properly approximated via DL algorithms, which approximate continuous functions. Cross-plots of type 2 seem to fix that issue by delivering higher $R^2$ factors and apparently nicer figures. However, they amplify the problem. In reality, the DL approximation of the inverse operator is inverting an incorrect forward approximation. Numerical results below illustrate this problem.

Obtaining high $R^2$ factors associated to cross-plots of type 3 (Equation~\ref{eq:crossplots}$_3$) is a challenging task as we discuss in Remark A of Section~\ref{loss_function}. Equation~\ref{exampleA_bad_result} shows a simple example in which cross-plots of type 1 and 2 deliver perfect $R^2$ marks and results, while cross-plots of type 3 are disastrous. This is also the situation that occurs in Example B.2. (see Table~\ref{tab:cp3_b1}). While the original training dataset is based on 1D Earth models, the one obtained after the predicted DNN inversion is a {\em piecewise} 1D Earth model, for which $\mathcal{F}_{\phi^\ast}$ is untrained for. When this occurs, the training database should be upgraded, either by increasing the space of the data samples or by selecting a different parameterization (e.g., measurements) for each sample. In our case, we choose to parametrize each sample independently (the later stategy) and we move to Example B.1. 

Table~\ref{tab:cp3_b1} shows mixed results for the Example B.1. Results without regularization are unremarkable with the geosignal forecasts showing poor results. The DNN inverse approximation accurately inverts for the outcome predicted by the DNN forward approximation. Nevertheless, since the DNN predicts solutions far from the true forward function, the predictions are poor. Again, this poor forecasting occurs because the DNN inverse approximation encounters subsurface models for which the forward DNN approximation is untrained. As a result, both the forward and inverse DDN approximations depart strongly from the true solutions. In other words, the inverse can only comply with their composition to be close to the identity, which is not robust to deliver accurate and physically relevant approximations.
\begin{table}[!h]
Cross-plots 3 \\
\begin{tabular}{ |l|c|c|c|c|c|c| } 
 \hline
                                              & Atten.          &  Atten. & Atten. & Phase & Phase  & Phase \\ 
     $R^2$ factors                  & LWD              &  Deep & Deep & LWD & Deep & Deep \\ 
                                             & Coaxial         & Coaxial &  Geosignal & Coaxial & Coaxial & Geosignal \\ 
 \hline
  Example B.1 & & & & & & \\
Without Reg.         &  0.9468         &  0.7406    &  0.0013              &   0.9383           &   0.9116      & 0.0167 \\ 
 With Reg.            &  0.9971          &  0.9979       & 0.9807               & 0.9969              & 0.9979         & 0.9856 \\ 
  \hline
  Example B.2 & & & & & & \\
Without Reg.         &  0.5721         &  0.8383    &  0.0253              &   0.4546           &   0.8611      & 0.0284 \\ 
With Reg.            &  0.9010          &  0.9701      & 0.5901               & 0.8621             & 0.9618         & 0.5877 \\ 
 \hline
\end{tabular}
\caption{\label{tab:cp3_b1}$R^2$ factors for Cross-plots 3 and Examples B.1 and B.2, with and without regularization, for the test dataset.}
\end{table}

To partially alleviate the above problem, we envision three possible solutions. First, we can increase the training dataset. This option is time-consuming and often impossible to achieve in practice. For example, herein, we already employ 1,000,000 samples. Second, we can include regularization. Results with regularization are of high quality (see Table~\ref{tab:cp3_b1}). However, the regularization term may hide alternative physical solutions of the inverse problem. Thus, the regularization diminishes the ability to perform uncertainty quantification. Similarly, it may induce on the user excessive confidence in the results. A third option is to consider the two-step loss function given by Equation~\ref{eq:loss4}. Following this approach, we first adjust the forward DNN approximation before training the DNN inverse approximation. Fixing the forward DNN often provides a proper forecast even in areas with a lower rate of training samples before producing a DNN approximation that approximates the inverse of the DNN forward approximation. Following this two-step approach without regularization, we obtain high $R^2$ factors for cross-plots of type 3: above 0.95 for the geosignal attenuation and phase, and above 0.99 for the remaining measurements.

\begin{figure}[!h]
\centering
	{%
	\includegraphics[scale=0.65]{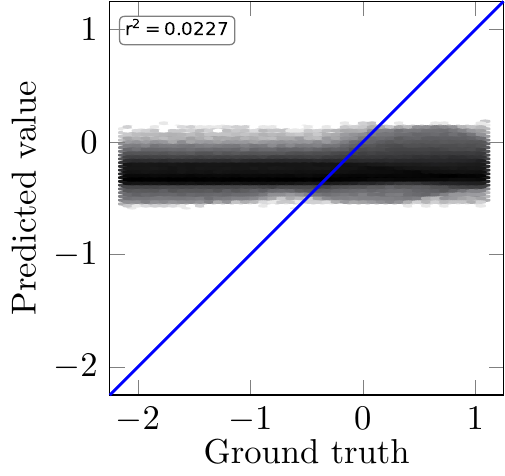}
	\hspace{0.2cm}
	\includegraphics[scale=0.65]{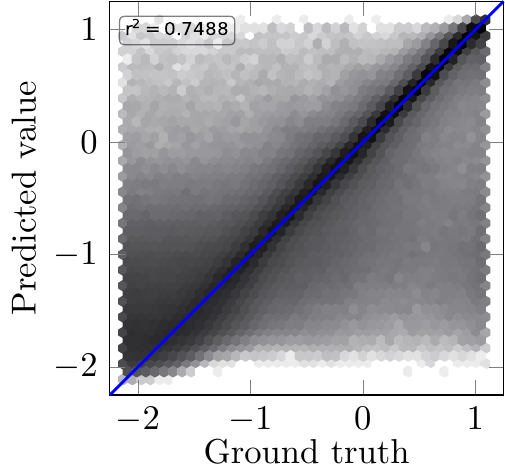}
	\hspace{0.2cm}
	\includegraphics[scale=0.65]{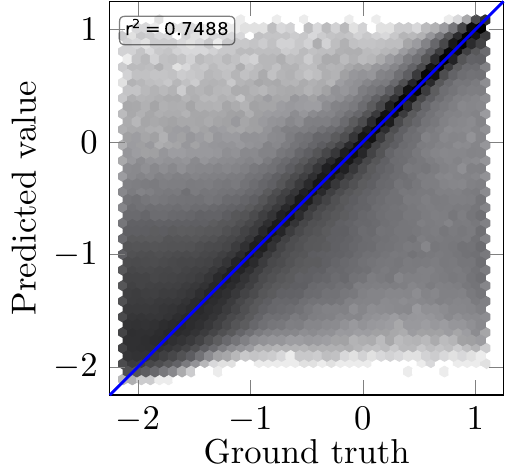}}
	{%
	\includegraphics[scale=0.65]{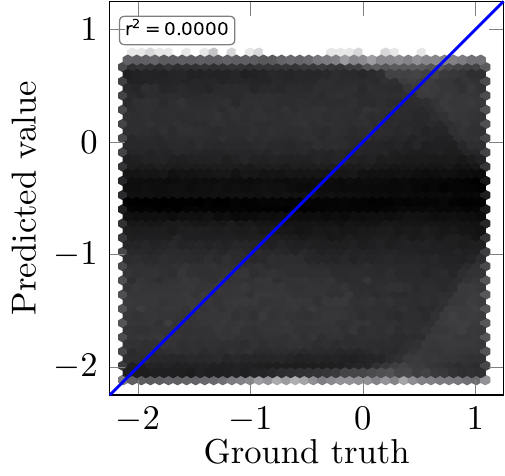}
	\hspace{0.2cm}
	\includegraphics[scale=0.65]{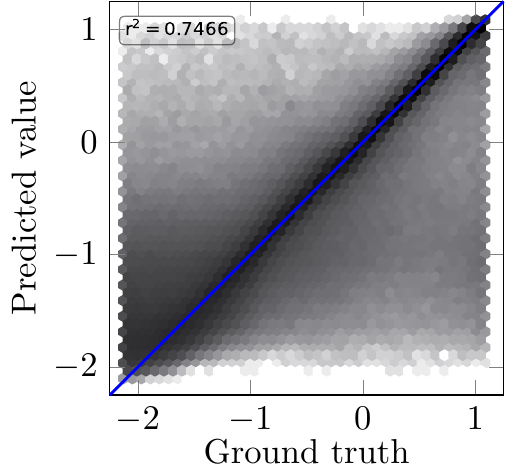}
	\hspace{0.2cm}
	\includegraphics[scale=0.65]{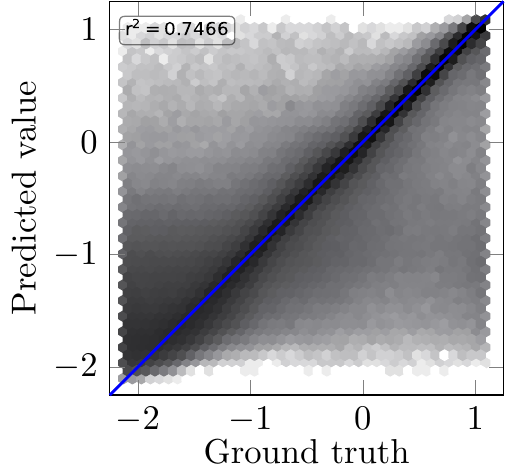}}
	{%
	\includegraphics[scale=0.65]{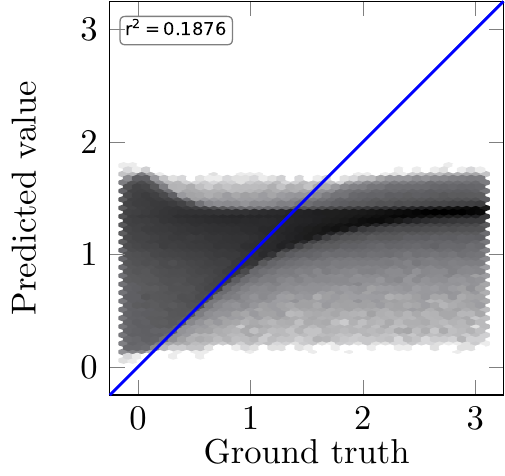}
	\hspace{0.2cm}
	\includegraphics[scale=0.65]{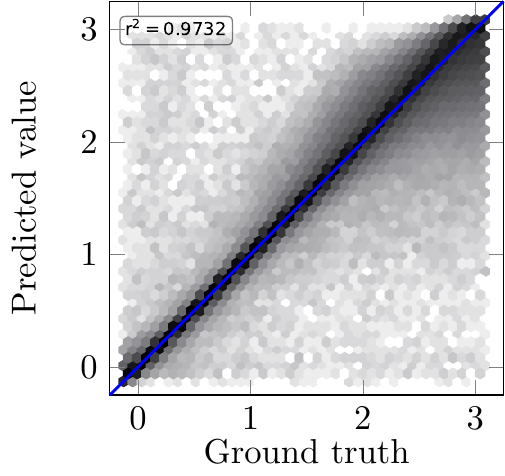}
	\hspace{0.2cm}
	\includegraphics[scale=0.65]{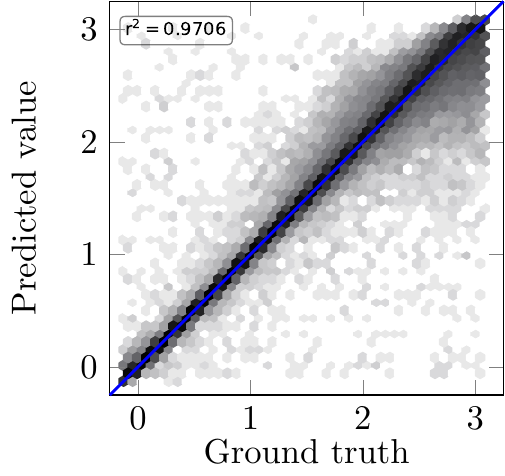}}
	{%
	\includegraphics[scale=0.65]{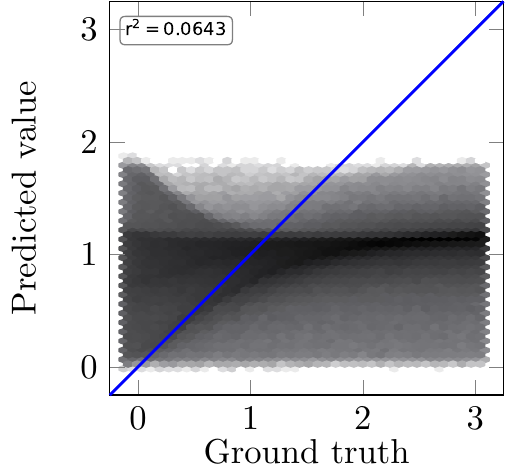}
	\hspace{0.2cm}
	\includegraphics[scale=0.65]{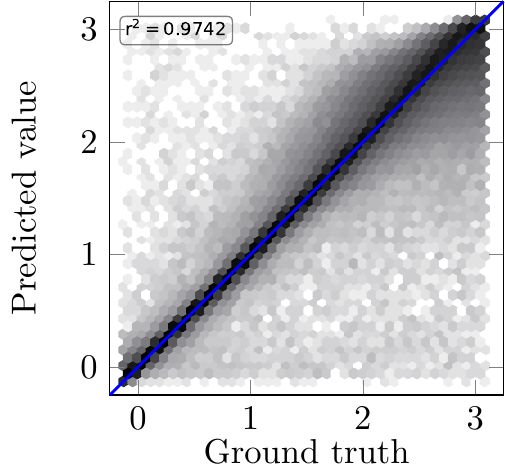}
	\hspace{0.2cm}
	\includegraphics[scale=0.65]{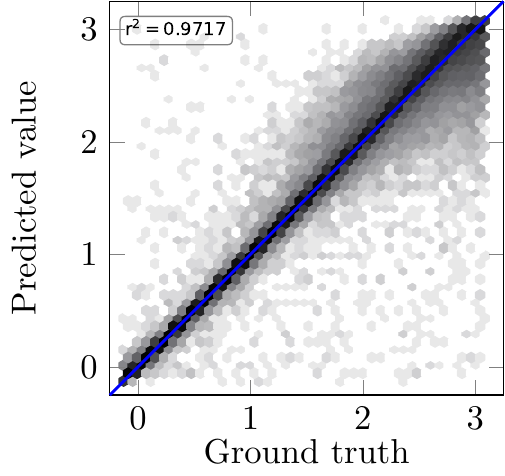}}
	{%
	\includegraphics[scale=0.65]{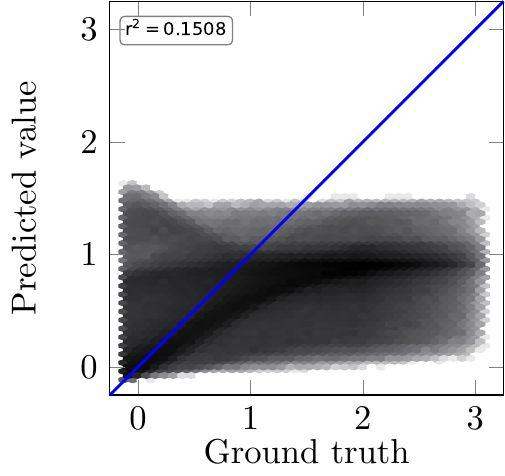}
	\hspace{0.2cm}
	\includegraphics[scale=0.65]{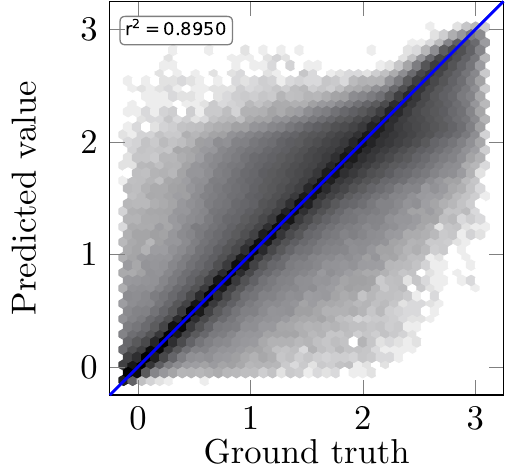}
	\hspace{0.2cm}
	\includegraphics[scale=0.65]{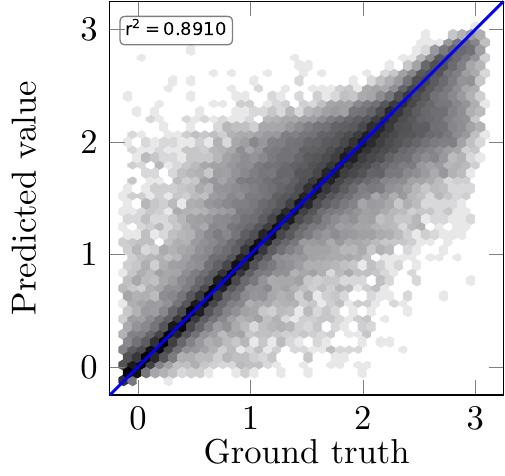}}
	\caption{\label{fig:cross-plots3}Cross-plots of type 4 for Example B.1 without regularization for the training dataset (first column), and with regularization for the training dataset (second column) and the test dataset (third column). First row: distance to the upper layer. Second row: distance to the lower layer. Third row: resistivity of upper layer. Fourth row: resistivity of lower layer. Fifth row: resistivity of central layer.}
\end{figure}

Finally, the $R^2$ factors for the cross-plots of type 4 do not reflect on the accuracy of the DNN algorithm, but rather on the nature of the inverse problem at hand. Low $R^2$ factors indicate there exist multiple solutions. A regularization term (e.g., Equation~\ref{eq:loss1}) increases the $R^2$ indicator. Figure~\ref{fig:cross-plots3} clearly illustrates this fact. However, it is misleading to conclude that results without regularization are always worse. They may simply exhibit a different (but still valid) solution of the inverse problem. 

\subsection{Inversion of realistic synthetic models}
	We now consider three realistic synthetic examples to assess the performance of the inversion process. In terms of log accuracy, we observe qualitatively similar results for the attenuation and phase logs. Thus, in the following we only display the attenuation logs and omit the phase logs.
	
\subsubsection{Model Problem I}
\begin{figure*}[!h]
		\centering
		\input{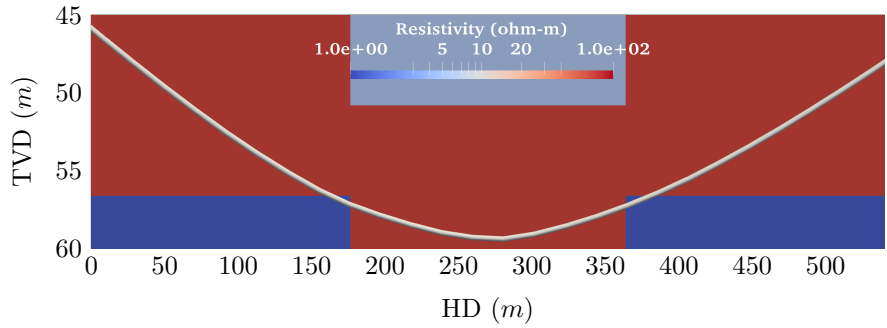}
	\caption{Formation of model problem I.}
	\label{fig:formation_model_1_original}
\end{figure*}

        Figure~\ref{fig:formation_model_1_original} describes a well trajectory in a synthetic model problem. The model has a resistive layer with a water-bearing layer underneath, and exhibits two geological faults.

        For the DNNs produced with the Example B.2 (with input measurements corresponding to 65 logging positions per sample), Figure~\ref{fig:formation_model_1_inverted_65} shows the corresponding inverted models using the Encoder-Decoder DNN with and without regularization. Results show inaccurate inversion results, specially for the case without regularization. Moreover, the predicted logs are far from the true logs, as Figure~\ref{fig:model_1_logs_65}, and as expected from cross-plots 3 (see Table~\ref{tab:cp3_b1}). The DNN inversion results are {\em piecewise} 1D models. However, the DNN approximation only trains with 1D models, not for piecewise 1D models, which explains the poor approximations they deliver (see Remark A on Section~\ref{loss_function}). 
\begin{figure*}[!h]
	\begin{subfigure}[]{\textwidth}
				\centering
	\input{./PAPER_ERROR_AND_LOSS/RESULTS_MOSTAFA/Syn_1/Predicted_F_FI_65.tex}%
	\caption{Without regularization}
\end{subfigure}
	\begin{subfigure}[]{\textwidth}
				\centering
	\input{./PAPER_ERROR_AND_LOSS/RESULTS_MOSTAFA/Syn_1/Predicted_F_FI_I_65.tex}%
	\caption{With regularization}
\end{subfigure}
	\caption{Inverted formation of model problem I using the inversion strategy of Example B.2, i.e., with input measurements corresponding to 65 logging positions per sample.}
	\label{fig:formation_model_1_inverted_65}
\end{figure*}
\begin{figure*}[!h]
	\begin{subfigure}[]{\textwidth}
		\centering
		\input{./PAPER_ERROR_AND_LOSS/RESULTS_MOSTAFA/Syn_1/Syn_1_FI_b/Atten-Syn-1-coaxial_65_without_regularization.tex}
						\captionsetup{belowskip=0.5em}  
		\caption{LWD coaxial measurement. Without regularization}
	\end{subfigure}
	
	\begin{subfigure}[]{\textwidth}
		\centering
		\input{./PAPER_ERROR_AND_LOSS/RESULTS_MOSTAFA/Syn_1/Syn_1_FI_b/Atten-Syn-1-coaxial_65.tex}
		\caption{LWD coaxial measurement. With regularization}
	\end{subfigure}
	\caption{Model problem I. Comparison between ${\cal F} \circ {\cal I}$ and ${\cal F} \circ {\cal I}_{\theta ^\ast}$ using the inversion strategy of Example B.2, i.e., with input measurements corresponding to 65 logging positions per sample.}
	\label{fig:model_1_logs_65}
\end{figure*}

In the remainder of this section, we restrict to DNNs produced with Example B.1. That is, we parametrize all observations at one location using information from that location alone. Figure~\ref{fig:formation_model_1_inverted_1} shows the corresponding inverted models. For the case of the Encoder-Decoder loss function without regularization, we observe in Figure~\ref{fig:formation_model_1_inverted_1}a an inverted model that is completely different from the original one. The corresponding logs (see Figure~\ref{fig:model_1_logs_without_reg}) are also inaccurate, as anticipated by the cross-plots results of type 3 shown in the previous subsection. When considering the two-step loss function without regularization, the recovered model (see Figure~\ref{fig:formation_model_1_inverted_1}b) is still quite different from the original one. Nonetheless, we observe a superb matching in the logs (see Figure~\ref{fig:model_1_logs_without_reg_two_step}), which indicates the presence of a different solution for the inverse problem. This confirms that the given measurements are insufficient to provide a unique solution for the inverse problem. For the case with regularization, inversion results (see Figure~\ref{fig:formation_model_1_inverted_1}b) match the original model, and the corresponding logs properly approximate the synthetic ones, see Figure~\ref{fig:model_1_logs_A}. Figures~\ref{fig:model_1_logs_B} and~\ref{fig:model_1_logs_C} confirm that our methodology delivers a proper training of the forward function approximation and the composition ${\cal F}_{\phi^\ast} \circ {\cal I}_{\theta ^\ast}$, respectively.
\begin{figure*}[!h]
	\begin{subfigure}[]{\textwidth}
				\centering
	\input{./PAPER_ERROR_AND_LOSS/RESULTS_MOSTAFA/Syn_1/Predicted_F_FI.tex}%
	\caption{Predicted formation using {\neww the Encoder-Decoder loss function} without regularization}
\end{subfigure}

	\begin{subfigure}[]{\textwidth}
	\centering
	\input{./PAPER_ERROR_AND_LOSS/RESULTS_MOSTAFA/Syn_1/Predicted_F_FI_two_step.tex}%
	\caption{Predicted formation using {\neww the two-step loss function} without regularization}
\end{subfigure}

	\begin{subfigure}[]{\textwidth}
				\centering
		\input{./PAPER_ERROR_AND_LOSS/RESULTS_MOSTAFA/Syn_1/predicted.tex}%
		\caption{Predicted formation using {\neww the Encoder-Decoder loss function} with regularization}
	\end{subfigure}
	\caption{Inverted formation of model problem I using the inversion strategy of Example B.1, i.e., with input measurements corresponding to one logging positions per sample.}
	\label{fig:formation_model_1_inverted_1}
\end{figure*}
\begin{figure*}[!h]
	\begin{subfigure}[]{\textwidth}
		\centering
		\input{./PAPER_ERROR_AND_LOSS/RESULTS_MOSTAFA/Syn_1/Syn_1_FI_b/Atten-Syn-1-coaxial_without_regularization.tex}
						\captionsetup{belowskip=0.5em}  
		\caption{LWD coaxial measurement}
	\end{subfigure}
	
	\begin{subfigure}[]{\textwidth}
		\centering
		\hspace*{-0.33 cm}\input{./PAPER_ERROR_AND_LOSS/RESULTS_MOSTAFA/Syn_1/Syn_1_FI_b/Atten-Syn-1_without_regularization.tex}
						\captionsetup{belowskip=0.5em}  
		\caption{Deep coaxial measurement}
	\end{subfigure}
	
	\begin{subfigure}[]{\textwidth}
		\centering
		\input{./PAPER_ERROR_AND_LOSS/RESULTS_MOSTAFA/Syn_1/Syn_1_FI_b/Atten-Syn-1-geo_without_regularization.tex}
		\caption{Geosignal measurement}
	\end{subfigure}
	\caption{Model problem I. Comparison between ${\cal F} \circ {\cal I}$ and ${\cal F} \circ {\cal I}_{\theta ^\ast}$ without regularization using the Encoder-Decoder loss function and the inversion strategy of Example B.1, i.e., with input measurements corresponding to one logging positions per sample.}
	\label{fig:model_1_logs_without_reg}
\end{figure*}

\begin{figure*}[!h]
	\begin{subfigure}[]{\textwidth}
		\centering
		\input{./PAPER_ERROR_AND_LOSS/RESULTS_MOSTAFA/Syn_1/Syn_1_FI_b/Atten-Syn-1-coaxial_two_step.tex}
		\captionsetup{belowskip=0.5em}  
		\caption{LWD coaxial measurement}
	\end{subfigure}
	
	\begin{subfigure}[]{\textwidth}
		\centering
		\hspace*{-0.33 cm}\input{./PAPER_ERROR_AND_LOSS/RESULTS_MOSTAFA/Syn_1/Syn_1_FI_b/Atten-Syn-1_two_step.tex}
		\captionsetup{belowskip=0.5em}  
		\caption{Deep coaxial measurement}
	\end{subfigure}
	
	\begin{subfigure}[]{\textwidth}
		\centering
		\input{./PAPER_ERROR_AND_LOSS/RESULTS_MOSTAFA/Syn_1/Syn_1_FI_b/Atten-Syn-1-geo_two_step.tex}
		\caption{Geosignal measurement}
	\end{subfigure}
	\caption{{\neww Model problem I. Comparison between ${\cal F} \circ {\cal I}$ and ${\cal F} \circ {\cal I}_{\theta ^\ast}$ using the two-step loss function without regularization and the inversion strategy of Example B.1, i.e., with input measurements corresponding to one logging positions per sample.}}
	\label{fig:model_1_logs_without_reg_two_step}
\end{figure*}

\begin{figure*}[!h]
	\begin{subfigure}[]{\textwidth}
		\centering
		\input{./PAPER_ERROR_AND_LOSS/RESULTS_MOSTAFA/Syn_1/Syn_1_FI_b/Atten-Syn-1-coaxial.tex}
						\captionsetup{belowskip=0.5em}  
		\caption{LWD coaxial measurement}
	\end{subfigure}
	
	\begin{subfigure}[]{\textwidth}
		\centering
		\hspace*{-0.33 cm}\input{./PAPER_ERROR_AND_LOSS/RESULTS_MOSTAFA/Syn_1/Syn_1_FI_b/Atten-Syn-1.tex}
						\captionsetup{belowskip=0.5em}  
		\caption{Deep coaxial measurement}
	\end{subfigure}
	
	\begin{subfigure}[]{\textwidth}
		\centering
		\input{./PAPER_ERROR_AND_LOSS/RESULTS_MOSTAFA/Syn_1/Syn_1_FI_b/Atten-Syn-1-geo.tex}
		\caption{Geosignal measurement}
	\end{subfigure}
	\caption{Model problem I. Comparison between ${\cal F} \circ {\cal I}$ and ${\cal F} \circ {\cal I}_{\theta ^\ast}$ with regularization using the inversion strategy of Example B.1, i.e., with input measurements corresponding to one logging positions per sample.}
	\label{fig:model_1_logs_A}
\end{figure*}
\begin{figure*}[!h]
	\begin{subfigure}[]{\textwidth}
		\centering
		\input{./PAPER_ERROR_AND_LOSS/RESULTS_MOSTAFA/Syn_1/Syn_1_F_aI/Atten-Syn-1-coaxial.tex}
				\captionsetup{aboveskip=0.2 em,belowskip=0.3em}  
		\caption{LWD coaxial measurement}
	\end{subfigure}

	\begin{subfigure}[]{\textwidth}
		\centering
		\hspace*{-0.6 cm}\input{./PAPER_ERROR_AND_LOSS/RESULTS_MOSTAFA/Syn_1/Syn_1_F_aI/Atten-Syn-1.tex}
		\captionsetup{belowskip=0.5em}  
		\caption{Deep coaxial measurement}
	\end{subfigure}
		
	\begin{subfigure}[]{\textwidth}
		\centering
		\input{./PAPER_ERROR_AND_LOSS/RESULTS_MOSTAFA/Syn_1/Syn_1_F_aI/Atten-Syn-1-geo.tex}
		\caption{Geosignal measurement}
	\end{subfigure}
	\caption{Model problem I. Comparison between ${\cal F} \circ {\cal I}$ and ${\cal F}_{\phi ^\ast} \circ {\cal I}$  with regularization using the inversion strategy of Example B.1, i.e., with input measurements corresponding to one logging positions per sample.}
	\label{fig:model_1_logs_B}
\end{figure*}
\begin{figure*}[!h]
	\begin{subfigure}[]{\textwidth}
		\centering
		\input{./PAPER_ERROR_AND_LOSS/RESULTS_MOSTAFA/Syn_1/Syn_1_F_aI_b//Atten-Syn-1-coaxial.tex}
						\captionsetup{aboveskip=0.2 em,belowskip=0.3em}  
		\caption{LWD coaxial measurement}
	\end{subfigure}

	\begin{subfigure}[]{\textwidth}
		\centering
		\hspace*{-0.6 cm}\input{./PAPER_ERROR_AND_LOSS/RESULTS_MOSTAFA/Syn_1/Syn_1_F_aI_b//Atten-Syn-1.tex}
				\captionsetup{belowskip=0.5em}  
		\caption{Deep coaxial measurement}
	\end{subfigure}
		
	\begin{subfigure}[]{\textwidth}
		\centering
		\input{./PAPER_ERROR_AND_LOSS/RESULTS_MOSTAFA/Syn_1/Syn_1_F_aI_b//Atten-Syn-1-geo.tex}
		\caption{Geosignal measurement}
	\end{subfigure}
	\caption{Model problem I. Comparison between ${\cal F} \circ {\cal I}$ and ${\cal F}_{\phi^\ast} \circ {\cal I}_{\theta ^\ast}$  with regularization using the inversion strategy of Example B.1, i.e., with input measurements corresponding to one logging positions per sample.}
	\label{fig:model_1_logs_C}
\end{figure*}

\subsubsection{Model Problem II}
	In this problem, we consider a 2.5m-thick conductive layer surrounded by two resistive layers. A well trajectory with a dip angle equal to 87$^\circ$ crosses the formation. Figure~\ref{fig:model_2} displays the original and predicted models by DL. This example illustrates some of the limitations of DNNs. In this case, the Earth models associated with part of the trajectory are outside the model problems considered in Section~\ref{problem_formulation}, which restrict to only one layer above and below the logging trajectory. Thus, the DNN is untrained for such models, and results cannot be trusted in those zones. Numerical results confirm these observations. Nonetheless, inaccurate inversion results are simple to identify  by inspection of the logs (Figures~\ref{fig:model_2_logs_A} and~\ref{fig:model_2_logs_B}). 
\begin{figure*}[!h]
	\begin{subfigure}[]{\textwidth}
\centering
		\input{./PAPER_ERROR_AND_LOSS/RESULTS_MOSTAFA/Syn_0/real.tex}
						\captionsetup{aboveskip=-0.9em,belowskip=0.5em}  
		\caption{Actual formation}
	\end{subfigure}
	\begin{subfigure}[]{\textwidth}
		\centering
		\input{./PAPER_ERROR_AND_LOSS/RESULTS_MOSTAFA/Syn_0/predicted.tex}%
		\caption{Predicted formation using one logging position with regularization}
	\end{subfigure}
\caption{Model problem 2. Comparison between actual and predicted formations with regularization using the inversion strategy of Example B.1, i.e., with input measurements corresponding to one logging positions per sample.}
	\label{fig:model_2}
\end{figure*}
\begin{figure*}[!h]
	\begin{subfigure}[]{\textwidth}
		\centering
		\input{./PAPER_ERROR_AND_LOSS/RESULTS_MOSTAFA/Syn_0/Syn_0_FI_b/Atten-Syn-1-coaxial.tex}
						\captionsetup{aboveskip=-0.9em,belowskip=0.5em}  
		\caption{LWD coaxial measurement}
	\end{subfigure}
	\begin{subfigure}[]{\textwidth}
		\centering
		\hspace*{-0.62 cm}\input{./PAPER_ERROR_AND_LOSS/RESULTS_MOSTAFA/Syn_0/Syn_0_FI_b/Atten-Syn-1.tex} 
						\captionsetup{aboveskip=-0.9em,belowskip=0.5em}  
		\caption{Deep coaxial measurement}
	\end{subfigure}
	\begin{subfigure}[]{\textwidth}
		\centering
		\input{./PAPER_ERROR_AND_LOSS/RESULTS_MOSTAFA/Syn_0/Syn_0_FI_b/Atten-Syn-1-geo.tex}
						\captionsetup{aboveskip=-0.9em,belowskip=0.5em}  
		\caption{Geosignal measurement}
	\end{subfigure}
	\caption{Model problem 2. Comparison between ${\cal F} \circ {\cal I}$ and ${\cal F} \circ {\cal I}_{\theta ^\ast}$ with regularization using the inversion strategy of Example B.1, i.e., with input measurements corresponding to one logging positions per sample.}
	\label{fig:model_2_logs_A}
\end{figure*}
\begin{figure*}[!h]
	\begin{subfigure}[]{\textwidth}
		\centering
		\input{./PAPER_ERROR_AND_LOSS/RESULTS_MOSTAFA/Syn_0/Syn_0_F_aI/Atten-Syn-1-coaxial.tex}
						\captionsetup{aboveskip=-0.9em,belowskip=0.5em}  
		\caption{LWD coaxial measurement}
	\end{subfigure}
	
	\begin{subfigure}[]{\textwidth}
		\centering
		\hspace*{-0.6 cm}\input{./PAPER_ERROR_AND_LOSS/RESULTS_MOSTAFA/Syn_0/Syn_0_F_aI/Atten-Syn-1.tex}
						\captionsetup{aboveskip=-0.9em,belowskip=0.5em}  
		\caption{Deep coaxial measurement}
	\end{subfigure}
	
	\begin{subfigure}[]{\textwidth}
		\centering
		\input{./PAPER_ERROR_AND_LOSS/RESULTS_MOSTAFA/Syn_0/Syn_0_F_aI/Atten-Syn-1-geo.tex}
						\captionsetup{aboveskip=-0.9em,belowskip=0.5em}  
		\caption{Geosignal measurement}
	\end{subfigure}
	\caption{Model problem 2. Comparison between ${\cal F} \circ {\cal I}$ and ${\cal F}_{\phi ^\ast} \circ {\cal I}$ with regularization using the inversion strategy of Example B.1, i.e., with input measurements corresponding to one logging positions per sample.}
	\label{fig:model_2_logs_B}
\end{figure*}

\subsubsection{Model Problem III}
We now consider a model formation exhibiting geological faults and two different well trajectories. For well trajectory 1, Figure~\ref{fig:model_3_trajectory_1} shows the model problem, logging trajectory, inversion results, and coaxial attenuation logs. Inversion results are excellent. When considering the second well trajectory shown in Figure~\ref{fig:model_3_trajectory_2}, we observe good inversion results except at the proximity of points with horizontal distance (HD) equals to 75m and 350m. These inaccurate inversion results are easily identified by examination of the corresponding logs.
\begin{figure*}[!h]
	\begin{subfigure}[]{\textwidth}
\centering
		\input{./PAPER_ERROR_AND_LOSS/RESULTS_MOSTAFA/Syn_2/real.tex}
		\caption{Actual formation}
	\end{subfigure}
	\begin{subfigure}[]{\textwidth}
		\centering
		\input{./PAPER_ERROR_AND_LOSS/RESULTS_MOSTAFA/Syn_2/predicted.tex}%
		\caption{Predicted formation}
	\end{subfigure}
	\begin{subfigure}[]{\textwidth}
		\centering
		\input{./PAPER_ERROR_AND_LOSS/RESULTS_MOSTAFA/Syn_2/Syn_2_FI_b/Atten-Syn-1-coaxial.tex}
		\caption{LWD coaxial measurement}
	\end{subfigure}
	\caption{Model problem III, trajectory 1. Comparison between actual and predicted formations and the corresponding coaxial logs with regularization using the inversion strategy of Example B.1, i.e., with input measurements corresponding to one logging positions per sample.}
	\label{fig:model_3_trajectory_1}
\end{figure*}
\begin{figure*}[!h]
	\begin{subfigure}[]{\textwidth}
		\centering
		\input{./PAPER_ERROR_AND_LOSS/RESULTS_MOSTAFA/Syn_3/real.tex}
		\caption{Actual formation}
	\end{subfigure}
	\begin{subfigure}[]{\textwidth}
		\centering
		\input{./PAPER_ERROR_AND_LOSS/RESULTS_MOSTAFA/Syn_3/predicted.tex}%
		\caption{Predicted formation}
	\end{subfigure}
	
		\begin{subfigure}[]{\textwidth}
		\centering
		\input{./PAPER_ERROR_AND_LOSS/RESULTS_MOSTAFA/Syn_3/Syn_3_FI_b/Atten-Syn-1-coaxial.tex}
		\caption{LWD coaxial measurement}
	\end{subfigure}
	\caption{Model problem III, Trajectory 2. Comparison between actual and predicted formations and the corresponding coaxial logs with regularization using the inversion strategy of Example B.1, i.e., with input measurements corresponding to one logging positions per sample.}
	\label{fig:model_3_trajectory_2}
\end{figure*}

	\section{Discussion and conclusions}\label{conclusions} 
In this work, we focus on the use of deep neural networks (DNN) for the inversion of borehole resistivity measurements for geosteering applications. We analyze the strong impact that different loss functions have on the prediction results. We illustrate via a simple benchmark example that a traditional data misfit loss function delivers poor results. As a remedy, we use an Encoder-Decoder or a two-step loss function. These approaches generate two DNN approximations: one for the forward function and another one for the inverse operator. We propose different neural network architectures for each approximation functions.

To guarantee that the inverse DNN approximation provides meaningful results, we need to ensure that the training dataset contains sufficient samples. Otherwise, both forward and inverse DNN operators may provide incorrect solutions while still ensuring the composition of both operators to be close to the identity. Thus, the approach is highly dependent on the existence of a sufficiently rich training dataset, which facilitates the learning process of the DNNs. In the case of 1D layered formations, it is often feasible to produce the required dataset. However, for more complicated cases, for example, the inversion of 2D and 3D geometries, a direct extension may be limited due to the larger number of inversion variables and the extremely time-consuming process of producing an exhaustive  dataset. 

As a partial remedy for this limitation, we find it highly beneficial to add a regularization term to the loss function based on the existing training dataset. This reduces the richness we need to guarantee within the training datasets. Nevertheless, such regularization terms may hide alternative feasible solutions for the inverse operator, which may provide excessive confidence on the results and minimize the capacity to perform a fair uncertainty quantification assessment. Another possibility to partially alleviate the aforementioned problem is to consider a two-step loss functions. Using this approach, we have shown that the inverse problem considered in this work admits different solutions that are physically feasible, a fact that was obscured when using the regularization term.

Other critical limitations of DNNs we encounter in this work are: (a) the limited approximation capabilities of DNNs to reproduce discontinuous functions, (b)  {\neww the need of a new dataset and trained DNN for each subsurface parametrization,} and (c) the poor results they exhibit when they are evaluated over a sample that is outside the training dataset space. More importantly, it is often difficult to identify the source of poor results, which may include inadequate selections of: (i) loss function, (ii) DNN architecture, (iii) regularization term, (iv) training dataset, (v) optimization algorithm, (vi) rescaling operator and norms, (vii) model parameterization, (viii) approximation capabilities of DNNs, or simply (ix) the nature of the problem due to a lack of adequate measurements.
To deal with the aforementioned limitations, we propose a careful step-by-step error control based on: (a) selecting adequate norms, (b) proper rescaling of the variables, (c) selecting a well suited loss function possibly with a regularization term, (d) analyzing the evolution of the different terms of the loss function, (e) studying multiple cross-plots of different nature, and (f) performing an in-depth assessment of the results over multiple realistic test examples.

Finally, we show it is possible to obtain a good-quality inversion of geosteering measurements with limited {\em online} computational cost, thus, suitable for real-time inversion. Moreover, the quality of the inversion results can be rapidly evaluated to detect its possible inaccuracies in the field and select alternative inversion methods when needed. 

Possible future research lines of this work include: 
(a) to study different DNN architectures when applied to these problems, for example, using automatic DNN architecture generators such as AutoML techniques, 
(b) to design proper measurement acquisition systems and adequate Earth model parametrizations using the cross-plots delivered by the DNNs, 
(c) to consider more complex Earth models, possibly containing geological faults or other relevant subsurface features, 
(d) to develop optimal sampling techniques for inverse problems, possibly containing a different number of samples to train the forward and inverse operators,
(e) to design and analyse new regularization techniques, 
(f) to use Bayesian DNNs for uncertainty quantification, 
{\neww and (g) to use transfer learning techniques for higher spatial dimensions, which can alleviate data requirements to train the corresponding DNNs}.
Finally, a natural step toward industrial applications is to evaluate the performance of our DNN approach when having noisy measurements. As mentioned above, we shall use our approach to design proper measurement acquisition techniques and adequate earth model parameterizations using the cross-plots delivered by the DNNs. 
	\section{Acknowledgments}
The research reported in this article has been funded by the European Union's Horizon 2020 research and innovation programme under the Marie Sklodowska-Curie grant agreement No 777778 (MATHROCKS), the European POCTEFA 2014-2020 Project PIXIL (EFA362/19) by the European Regional Development Fund (ERDF) through the Interreg V-A Spain-France-Andorra programme, the Austrian Ministry for Transport, Innovation and Technology (BMVIT), the Federal Ministry for Digital and Economic Affairs (BMDW), the Province of Upper Austria in the frame of the COMET - Competence Centers for Excellent Technologies Program managed by Austrian Research Promotion Agency FFG, the Project of the Spanish Ministry of Economy and Competitiveness with reference MTM2016-76329-R (AEI/FEDER, EU), the BCAM ``Severo Ochoa'' accreditation of excellence (SEV-2017-0718), and the Basque Government through the BERC 2018-2021 program, the two Elkartek projects ArgIA (KK-2019-00068) and MATHEO (KK-2019-00085), the Consolidated Research Group MATHMODE (IT1294-19) given by the Department of Education, The University of Texas at Austin Research Consortium on Formation Evaluation, jointly sponsored by Anadarko, Aramco, Baker Hughes, BHP, BP, Chevron, China Oilfield Services Limited, CNOOC International, ConocoPhillips, DEA, Eni, Equinor ASA, ExxonMobil, Halliburton, INPEX, Lundin Norway, Occidental, Oil Search, Petrobras, Repsol, Schlumberger, Shell, Southwestern, Total, Wintershall Dea, and Woodside Petroleum Limited. Carlos Torres-Verd\'in is grateful for the financial support provided by the Brian James Jennings Memorial Endowed Chair in Petroleum and Geosystems Engineering. This publication acknowledges the financial support of the CSIRO Professorial Chair in Computational Geoscience at Curtin University and the Deep Earth Imaging Enterprise Future Science Platforms of the Commonwealth Scientific Industrial Research Organisation, CSIRO, of Australia.  Additionally, at Curtin University, The Institute for Geoscience Research (TIGeR) and by the Curtin Institute for Computation, kindly provide continuing support.
	
	\bibliography{PAPER_ERROR_AND_LOSS/mybibfile}

\begin{thebibliography}{10}
\expandafter\ifx\csname url\endcsname\relax
  \def\url#1{\texttt{#1}}\fi
\expandafter\ifx\csname urlprefix\endcsname\relax\def\urlprefix{URL }\fi
\expandafter\ifx\csname href\endcsname\relax
  \def\href#1#2{#2} \def\path#1{#1}\fi

\bibitem{Lu}
L.~Lu, Y.~Zheng, G.~Carneiro, L.~Yang, Deep Learning for Computer Vision:
  Expert techniques to train advanced neural networks using TensorFlow and
  Keras, Springer, Switzerland, 2017.

\bibitem{Yu}
D.~Yu, L.~Deng, Automatic Speech Recognition: A Deep Learning approach,
  Springer, London, 2017.

\bibitem{Kumar}
B.~Bhanu, A.~Kumar, Deep Learning for Biometrics, Springer, Switzerland, 2017.

\bibitem{He}
K.~He, X.~Zhang, S.~Ren, J.~Sun, Deep residual learning for image recognition,
  arXiv:1512.03385 (2015).

\bibitem{Badrinarayanan}
V.~Badrinarayanan, A.~Kendall, R.~Cipolla, Segnet: A deep convolutional
  encoder-decoder architecture for image segmentation, IEEE Transactions on
  Pattern Analysis and Machine Intelligence 39~(12) (2017) 2481--2495.

\bibitem{chollet2015}
F.~Chollet, Keras, \url{https://github.com/fchollet/keras} (2015).

\bibitem{paszke2017automatic}
A.~Paszke, S.~Gross, S.~Chintala, G.~Chanan, E.~Yang, Z.~DeVito, Z.~Lin,
  A.~Desmaison, L.~Antiga, A.~Lerer, Automatic differentiation in pytorch, in:
  NIPS-W, 2017.

\bibitem{ZHAO2019213}
R.~Zhao, R.~Yan, Z.~Chen, K.~Mao, P.~Wang, R.~X. Gao, Deep learning and its
  applications to machine health monitoring, Mechanical Systems and Signal
  Processing 115 (2019) 213 -- 237.
\newblock \href {https://doi.org/https://doi.org/10.1016/j.ymssp.2018.05.050}
  {\path{doi:https://doi.org/10.1016/j.ymssp.2018.05.050}}.

\bibitem{6639345}
L.~{Deng}, J.~{Li}, J.~{Huang}, K.~{Yao}, D.~{Yu}, F.~{Seide}, M.~{Seltzer},
  G.~{Zweig}, X.~{He}, J.~{Williams}, Y.~{Gong}, A.~{Acero}, Recent advances in
  deep learning for speech research at microsoft, in: 2013 IEEE International
  Conference on Acoustics, Speech and Signal Processing, 2013, pp. 8604--8608.

\bibitem{LANGKVIST201411}
M.~Längkvist, L.~Karlsson, A.~Loutfi, A review of unsupervised feature
  learning and deep learning for time-series modeling, Pattern Recognition
  Letters 42 (2014) 11 -- 24.
\newblock \href {https://doi.org/https://doi.org/10.1016/j.patrec.2014.01.008}
  {\path{doi:https://doi.org/10.1016/j.patrec.2014.01.008}}.

\bibitem{quteprints127354}
R.~Vargas, A.~Mosavi, R.~Ruiz, Deep learning: A review, Working paper (January
  2017).

\bibitem{JAN2019275}
B.~Jan, H.~Farman, M.~Khan, M.~Imran, I.~U. Islam, A.~Ahmad, S.~Ali, G.~Jeon,
  Deep learning in big data analytics: A comparative study, Computers \&
  Electrical Engineering 75 (2019) 275 -- 287.
\newblock \href
  {https://doi.org/https://doi.org/10.1016/j.compeleceng.2017.12.009}
  {\path{doi:https://doi.org/10.1016/j.compeleceng.2017.12.009}}.

\bibitem{Shahriari_deep_inverse}
M.~Shahriari, D.~Pardo, A.~Pic{\'{o}}n, A.~Galdran, J.~D. Ser,
  C.~Torres{-}Verd{\'{\i}}n, A deep learning approach to the inversion of
  borehole resistivity measurements, Computational Geosciences (2020).
\newblock \href {https://doi.org/10.1007/s10596-019-09859-y}
  {\path{doi:10.1007/s10596-019-09859-y}}.

\bibitem{Pardo}
D.~Pardo, C.~Torres-Verdin, Fast 1{D} inversion of logging-while-drilling
  resistivity measurements for the improved estimation of formation resistivity
  in high-angle and horizontal wells, Geophysics 80 (2) (2014) E111--E124.

\bibitem{Ijasan}
O.~Ijasana, C.~Torres-Verd\'{i}n, W.~E. Preeg, Inversion-based petrophysical
  interpretation of logging-while-drilling nuclear and resistivity
  measurements, Geophysics 78 (6) (2013) D473--D489.

\bibitem{Desbrandes}
R.~Desbrandes, R.~Clayton, Chapter 9 measurement while drilling, Developments
  in Petroleum Science 38 (1994) 251 -- 279.

\bibitem{Ghasemi}
M.~Ghasemi, Y.~Yang, E.~Gildin, Y.~Efendiev, V.~M. Calo, Fast multiscale
  reservoir simulations using pod-deim model reduction, Society of Petroleum
  Engineers (2015) 1--18.

\bibitem{Chemali}
R.~Chemali, M.~Bittar, F.~Hveding, M.~Wu, M.~Dautel, Improved geosteering by
  integrating in real time images from multiple depths of investigation and
  inversion of azimuthal resistivity signals, Society of Petrophysicists and
  Well-Log Analysts (2010) 1--7.

\bibitem{Dupuis}
C.~Dupuis, J.~M. Denichou, Automatic inversion of deep-directional-resistivity
  measurements for well placement and reservoir description, The Leading Edge
  34~(5) (2015) 504--512.

\bibitem{Zhang}
Z.~Zhang, N.~Yuan, C.~R. Liu, 1-{D} inversion of triaxial induction logging in
  layered anisotropic formation, Progress In Electromagnetics Research B 44
  (2012) 383--403.

\bibitem{Seifert}
D.~J. Seifert, S.~A. Dossary, R.~E. Chemali, M.~S. Bittar, A.~A. Lotfy, J.~L.
  Pitcher, M.~A. Bayrakdar, Deep electrical images, geosignal, and real-time
  inversion help guide steering decisions, Society of Petroleum Engineers
  (2009) 1--9.

\bibitem{Tarantola}
A.~Tarantola, Inverse Problem Theory and Methods for Model Parameter
  Estimation, Society for Industrial and Applied Mathematics, 2005.

\bibitem{Shahriari}
M.~Shahriari, S.~Rojas, D.~Pardo, A.~Rodr\'{i}guez-Rozas, S.~A. Bakr, V.~M.
  Calo, I.~Muga, A numerical 1.5{D} method for the rapid simulation of
  geophysical resistivity measurements, Geosciences 8(6) (2018) 1--28.

\bibitem{8746808}
L.~{Wang}, H.~{Li}, Y.~{Fan}, Bayesian inversion of logging-while-drilling
  extra-deep directional resistivity measurements using parallel tempering
  markov chain monte carlo sampling, IEEE Transactions on Geoscience and Remote
  Sensing 57~(10) (2019) 8026--8036.

\bibitem{Malinverno_2000}
A.~Malinverno, C.~Torres-Verd{\'{\i}}n, Bayesian inversion of {DC} electrical
  measurements with uncertainties for reservoir monitoring, Inverse Problems
  16~(5) (2000) 1343--1356.
\newblock \href {https://doi.org/10.1088/0266-5611/16/5/313}
  {\path{doi:10.1088/0266-5611/16/5/313}}.

\bibitem{doi:10.1190/1.2713043}
J.~Gunning, M.~E. Glinsky, Detection of reservoir quality using bayesian
  seismic inversion, GEOPHYSICS 72~(3) (2007) R37--R49.
\newblock \href {https://doi.org/10.1190/1.2713043}
  {\path{doi:10.1190/1.2713043}}.

\bibitem{Vogel}
C.~Vogel, Computational Methods for Inverse Problems, Society for Industrial
  and Applied Mathematics, 2002.

\bibitem{Higham}
C.~F. Higham, D.~J. Higham, Deep learning: An introduction for applied
  mathematicians, Computing Research Repository abs/1801.05894 (2018).

\bibitem{Shahriari_deep_forward}
M.~Shahriari, D.~Pardo, B.~Moser, F.~Sobieczky, A deep neural network as
  surrogate model for forward simulation of borehole resistivity measurements,
  Procedia Manufacturing 42 (2020) 235 -- 238, international Conference on
  Industry 4.0 and Smart Manufacturing (ISM 2019).

\bibitem{Aramberri}
J.~Alvarez-Aramberri, D.~Pardo, Dimensionally adaptive hp-finite element
  simulation and inversion of 2{D} magnetotelluric measurements, Journal of
  Computational Science 18 (2017) 95--105.

\bibitem{Bakr}
S.~A. Bakr, D.~Pardo, T.~Mannseth, Domain decomposition {F}ourier {FE} method
  for the simulation of 3{D} marine {CSEM} measurements, J. Comput. Phys. 255
  (2013) 456--470.

\bibitem{Davydycheva}
S.~Davydycheva, T.~Wang, A fast modelling method to solve {M}axwell's equations
  in 1{D} layered biaxial anisotropic medium, Geophysics 76 (5) (2011)
  F293--F302.

\bibitem{Davydycheva1}
S.~Davydycheva, D.~Homan, G.~Minerbo, Triaxial induction tool with electrode
  sleeve: {FD} modeling in 3{D} geometries, Journal of Applied Geophysics 67
  (2004) 98--108.

\bibitem{Puzyrev}
V.~Puzyrev, {Deep learning electromagnetic inversion with convolutional neural
  networks}, Geophysical Journal International 218~(2) (2019) 817--832.

\bibitem{Moghadas}
D.~Moghadas, {One-dimensional deep learning inversion of electromagnetic
  induction data using convolutional neural network}, Geophysical Journal
  International 222~(1) (2020) 247--259.

\bibitem{Loseth}
L.~O. Loseth, B.~Ursin, Electromagnetic fields in planarly layered anisotropic
  media, Geophysical Journal International 170 (2007) 44--80.

\bibitem{rahimpour2017feature}
A.~Rahimpour, A.~Taalimi, H.~Qi, Feature encoding in band-limited distributed
  surveillance systems, in: 2017 IEEE International Conference on Acoustics,
  Speech and Signal Processing (ICASSP), IEEE, 2017, pp. 1752--1756.

\bibitem{jin2019using}
Y.~Jin, X.~Wu, J.~Chen, Y.~Huang, et~al., Using a physics-driven deep neural
  network to solve inverse problems for lwd azimuthal resistivity measurements,
  in: SPWLA 60th Annual Logging Symposium, Society of Petrophysicists and
  Well-Log Analysts, 2019.

\bibitem{tf2}
M.~Abadi, A.~Agarwal, P.~Barham, E.~Brevdo, Z.~Chen, C.~Citro, G.~S. Corrado,
  A.~Davis, J.~Dean, M.~Devin, S.~Ghemawat, I.~Goodfellow, A.~Harp, G.~Irving,
  M.~Isard, Y.~Jia, R.~Jozefowicz, L.~Kaiser, M.~Kudlur, J.~Levenberg,
  D.~Man\'{e}, R.~Monga, S.~Moore, D.~Murray, C.~Olah, M.~Schuster, J.~Shlens,
  B.~Steiner, I.~Sutskever, K.~Talwar, P.~Tucker, V.~Vanhoucke, V.~Vasudevan,
  F.~Vi\'{e}gas, O.~Vinyals, P.~Warden, M.~Wattenberg, M.~Wicke, Y.~Yu,
  X.~Zheng, {TensorFlow}: Large-scale machine learning on heterogeneous
  systems, software available from tensorflow.org (2015).

\bibitem{quan2016fusionnet}
T.~M. Quan, D.~G.~C. Hildebrand, W.-K. Jeong, Fusionnet: A deep fully residual
  convolutional neural network for image segmentation in connectomics (2016).
\newblock \href {http://arxiv.org/abs/1612.05360} {\path{arXiv:1612.05360}}.

\bibitem{NIPS2012_4824}
A.~Krizhevsky, I.~Sutskever, G.~E. Hinton, Imagenet classification with deep
  convolutional neural networks, in: F.~Pereira, C.~J.~C. Burges, L.~Bottou,
  K.~Q. Weinberger (Eds.), Advances in Neural Information Processing Systems
  25, Curran Associates, Inc., 2012, pp. 1097--1105.

\bibitem{7949028}
K.~H. {Jin}, M.~T. {McCann}, E.~{Froustey}, M.~{Unser}, Deep convolutional
  neural network for inverse problems in imaging, IEEE Transactions on Image
  Processing 26~(9) (2017) 4509--4522.

\end{thebibliography}
        \begin{appendices}
	\section{Measurements acquisition system \label{app:meas}}
	
	We first consider a co-axial LWD instrument equipped with two transmitters and two receivers (see Figure \ref{fig_device_lwd}). ${H}^1_{zz}$ and ${H}^2_{zz}$ are the $zz$-couplings of the magnetic field measured at the first and the second receivers, respectively (the first and second subscripts denote the orientation of the transmitter and receiver, respectively). Then, we define the attenuation and phase difference as follows:
	\begin{figure*}[ht]
	\centering
	\begin{tikzpicture}[scale=1.0]
	\node (layers) at (0,0)[scale=1]{
		\input{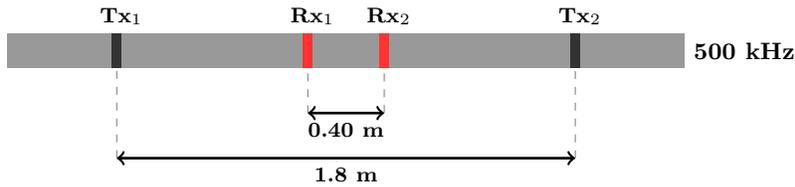} 
	};
	\end{tikzpicture}
	\caption{Conventional LWD logging instrument. Tx$_i$ and Rx$_i$ are the transmitters and the receivers, respectively.}
	\label{fig_device_lwd}
\end{figure*}

	\begin{equation}
	\begin{split}
	\ln \frac{{H}^1_{zz}}{{H}^2_{zz}}= \underbrace{\ln \frac{\mid {H}^1_{zz}\mid}{\mid{H}^2_{zz}\mid}}_{\times 20 \log(e) =: \text{attenuation }(dB)}+i \underbrace{\left( ph({H}^1_{zz})-ph({H}^2_{zz})\right)}_{\times \dfrac{180}{\pi} =:\text{phase difference (degree)}},
	\end{split}
	\label{eq:atten_phase}
	\end{equation}
	where $ph$ denotes the phase of a complex number. We then record the average of the attenuations and phase differences associated with the two transmitters, and we denote these values as {\em LWD coaxial}. 
	
\begin{figure*}[ht]
		\centering
		\begin{tikzpicture}[scale=1.0]
		\node (layers) at (0,0)[scale=1]{
			\input{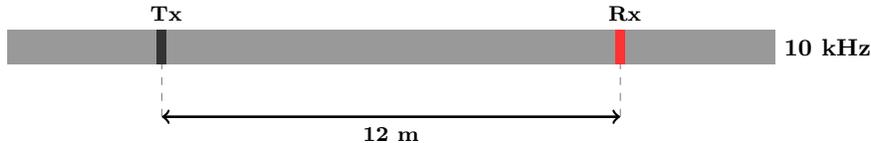} 
		};
		\end{tikzpicture}
		\caption{Short-spacing of a deep azimuthal logging instrument. Tx and Rx are the transmitter and the receiver, respectively.}
		\label{fig_device_azim}
	\end{figure*}

\begin{figure}
	\centering
	\input{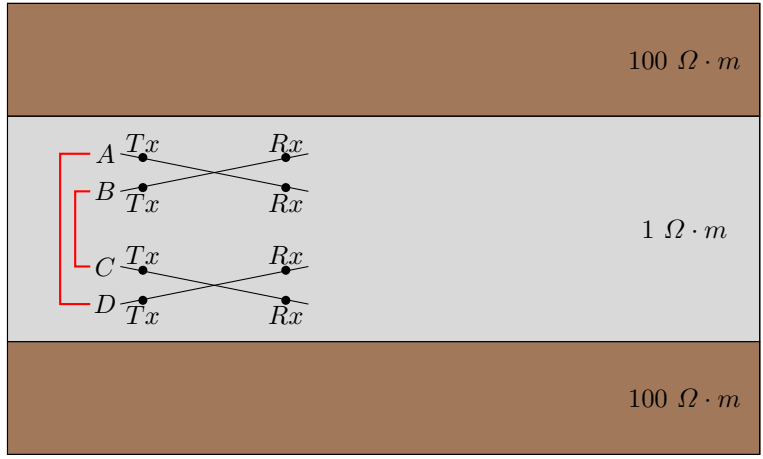}
	\caption{Illustration with four logging trajectories. By symmetry, measurements recorded with trajectories A and D are identical. The same occurs with trajectories B and C. If these measurements are continuous with respect to the dip angle, then at 90 degrees they all become identical, which disables the possibility of identifying if a nearby bed boundary is located on top or on the bottom of the logging instrument.}
	\label{fig:1D_4_trajectories}
\end{figure}

	Then, we consider a short-spacing configuration corresponding to a deep azimuthal instrument equipped with one transmitter and one receiver, as shown in Figure \ref{fig_device_azim}. 
	In this logging instrument, the distance between transmitter and receiver is significantly larger than that of the previously considered LWD instrument. It also employs tilted receivers that are sensitive to the presence of bed boundaries. We record several measurements with this logging instrument: (a) the attenuation and phase differences, denotes as {\em deep coaxial}, computed using Equation \eqref{eq:atten_phase} with ${H}^2_{zz}=1$, and (b) the attenuation and phase differences of a directional measurement expressed as:
	\begin{equation}
	\begin{split}
	Geosignal=\ln \frac{{H}_{zz}-{H}_{zx}}{{H}_{zz}+{H}_{zx}}&= \underbrace{\ln \frac{\mid {H}_{zz}-{H}_{zx}\mid}{\mid{H}_{zz}+{H}_{zx}\mid}}_{\times 20 \log(e) =: \text{attenuation }(dB)}\\
	&+i \underbrace{\left( ph({H}_{zz}-{H}_{zx})-ph({H}_{zz}+{H}_{zx})\right)}_{\times \dfrac{180}{\pi} =:\text{phase difference (degree)}}.
	\end{split}
	\end{equation}
We denote it as {\em geosignal}. These measurements exhibit a discontinuity as a function of the dip angle at 90 degrees. Indeed, such discontinuity is essential in the measurements if one wants to discern between top and bottom of the logging instrument---see Figure \ref{fig:1D_4_trajectories}.   

We consider two types of logging trajectories. In one case, corresponding to Example B.1, each trajectory consists of a single logging position, and the set of measurements is given by 6 real numbers. For the Example B.2, each trajectory contains 65 logging positions with a logging step size of $0.3048\ m$. Thus, our set of measurements per logging trajectory consist of 195 pairs of real numbers (see \cite{Shahriari_deep_inverse, Shahriari_deep_forward} for further details).

	\section{DDN Arquitectures \label{sec:dnn_arquitectures}}

In this work, we use DNN architectures based on residual blocks \cite{He,quan2016fusionnet} with convolutional operators \cite{Higham,NIPS2012_4824,7949028} to approximate the forward and the inverse problems.
\subsection{Forward Problem DNN Architecture}
We consider:
\begin{equation}
\mathcal{F}_{{\cal R},\phi} := \mathbf{l}^{c_1}_{\phi_{N}} \circ \mathcal{L}  \circ \mathcal{B}^{N-1}_{\phi_{N-1}} \circ \mathcal{B}^{N-2}_{\phi_{N-2}} \circ \cdots \circ \mathcal{B}^{1}_{\phi_{1}},
\label{dnn_function_forward}
\end{equation}
where ${\phi}=\{{\phi}_i : \ i=1,\cdots, N\}$ is a set of all weights associated to each block and layer. In Equation \eqref{dnn_function_forward}, $\{\mathcal{B}^{i}_{\phi_{i}}: i=1,\cdots,5\}$ consist of an upsampling layer followed by residual blocks. Specifically:
\begin{equation}
\mathcal{B}^i_{\phi_{i}}  :=
\left({\cal N} \circ \mathbf{l}^{c_i}_{\phi^{1}_{i}}  \circ {\cal N} \circ \mathbf{l}^{c_i}_{\phi^{2}_{i}} + \mathcal{I} \right) \circ \mathcal{U},
\label{res_block}
\end{equation}
where $\mathbf{l}^{c}_{\phi}$ is a convolutional layer with $c_i$ as its convolution window, $\mathcal{I}$ is the identity function, $\mathcal{U}$ is a one-dimensional upsampling  that raises the dimension of the output of the residual block gradually to avoid missing information due to a sudden expansion of dimension, $\mathbf{l}^{c_1}_{\phi_{N}}$ is  the final convolutional layer  that acts as the final feature extractor, and ${\cal N}$ is a nonlinear \textit{activation function} \cite{Higham}. In our case, we select  ${\cal N}$ as the \textit{rectified linear unit (ReLU)}, which is defined as follows:
\begin{equation}
{\cal N}(x_1,x_2,\cdots,x_n) = \left( \max(0,x_1),\max(0,x_2), \cdots, \max(0,x_n)\right).
\end{equation}
We send the output of the final residual block to a bilinear upsampling $\mathcal{L}$ to expand the output dimension.

\subsection{Inverse Problem DNN Architecture}
Analogously as for the forward problem, we consider the following architecture:
\begin{equation}
\mathcal{F}_{{\cal R},\theta} :=  \mathcal{C}_{\theta_{N}} \circ \mathcal{S}  \circ \mathcal{B}^{N-1}_{\theta_{N-1}} \circ \mathcal{B}^{N-2}_{\theta_{N-2}} \circ \cdots \circ \mathcal{B}^{1}_{\theta_{1}},
\label{dnn_function_inverse}
\end{equation}
where ${\theta}=\{{\theta}_i : \ i=1,\cdots, N\}$ is a set of all the weights associated to each block and layer. In Equation \eqref{dnn_function_inverse}, the residual blocks $\{\mathcal{B}^{i}_{\theta_{i}}: i=1,\cdots,6\}$ are analogous to the ones in Equation \eqref{res_block}, excluding the one dimensional upsampling. In this architecture, the convolutional layers perform the down-sampling. $\mathcal{S}$ is a reshaping layer and $\mathcal{C}_{\theta_{N}}$ is a fully-connected layer, which performs the ultimate feature extraction and further down-sampling.

Using the above DNN architectures, minimization problems of Section \ref{loss_function} are optimized end-to-end with respect to $\phi$ and $\theta$ to obtain $\phi^\ast$ and $\theta^\ast$. 

	\end{appendices} 

\end{document}